\documentclass[%
 aip,
 amsmath,amssymb,
 reprint,%
]{revtex4-1}

\usepackage{graphicx}
\usepackage{dcolumn}
\usepackage{bm}

\usepackage[utf8]{inputenc}
\usepackage[T1]{fontenc}
\usepackage{mathptmx}
\usepackage{etoolbox}
\usepackage{natbib}
\usepackage{subfig}
\usepackage{amssymb}
\usepackage{comment}
\usepackage{subcaption}
\DeclareMathOperator{\sech}{sech}
\setcitestyle{numbers,square} 

\makeatletter
\def\@email#1#2{%
 \endgroup
 \patchcmd{\titleblock@produce}
  {\frontmatter@RRAPformat}
  {\frontmatter@RRAPformat{\produce@RRAP{*#1\href{mailto:#2}{#2}}}\frontmatter@RRAPformat}
  {}{}
}%
\makeatother
\begin{document}


\title{Bailout Embedding and Stability Analysis of a Dynamical Mean-Field Ising Model of Opinion Dynamics}

\author{Senbagaraman Sudarsanam}%
 \email{sraman@fmitech.net.}
\affiliation{ 
FMI Technologies LLC, New York, United States
}%

\date{\today}

\begin{abstract}
We study the stability of a discrete-time dynamical mean-field Ising model to perturbations. This model belongs to a broader class of models often used in the study of opinion dynamics in financial markets. In the presence of noise, these iterated maps are known to exhibit dynamics which resemble empirically observed behavior of financial markets, such as volatility clustering. Research in the recent past has identified attractor bubbling as one of the underlying mechanisms that lead to clustering of high volatility events in models of opinion dynamics. In this work, we employ the method of bailout embedding to create an extended, higher-dimensional system of iterated maps where the bubbling dynamics is isolated from dynamics in the original lower-dimensional space. The bailout embedding technique also introduces a bailout parameter and an associated bailout function which allows us to systematically induce several regimes of stability in the opinion dynamics and thus enables us to better understand the onset of attractor bubbling in this system. We demonstrate the occurrence of intermittency in opinion dynamics under appropriate conditions, with the discrete opinion dynamics sharing characteristics of empirically observed financial time series with volatility clustering. Inspired by the original context under which bailout embeddings were introduced, we propose an interpretation of the bailout function as a measure of investor inertia in the markets. This work hopes to demonstrate the usefulness of bailout embeddings as a tool to study mathematical models of financial market dynamics.
\end{abstract}

\maketitle

\begin{quotation}
There is mounting evidence suggesting that one of the mechanisms that lead to volatility clustering and power law distribution of returns in financial markets resembles attractor bubbling and on-off intermittency in dynamical systems. A physical system that is known to exhibit attractor bubbling and on-off intermittency is the dynamics of finite sized particles in fluid flows which is modeled using the Maxey-Riley equation. Bailout embedding is an analytical embedding technique inspired by the dynamics of finite sized particles and is useful for separating out bubbling dynamics from the underlying dynamics on invariant manifolds with locally unstable subsets. We use this framework to probe a discrete-time mean-field model of interacting agents in financial markets and find that the bailout function can be varied to induce various levels of stability in the underlying dynamics. We also demonstrate that the model of opinion dynamics chosen for this study exhibits intermittency and power law behavior upon perturbation under appropriate conditions, similar to empirically observed financial time series data. We hope to demonstrate the utility of bailout embeddings as an effective tool for the study of volatility clustering and other extreme events in models of financial markets.
\end{quotation}

\section{Introduction}
The Ising model is a well-known mathematical model that is most often used in statistical physics but has also found extensive application in models of social interactions\citep{McCoy1973, Chowdhury1999,Kaizoji2000, Hołyst2000, Sornette2014}. The collective evolution of opinions among market participants, who each hold individual positions that evolve through interactions with their environment and external ordering forces such as news, is crucial to the understanding of financial market dynamics\citep{Quanbo2020}. An empirically observed market phenomenon that has received a lot of attention is the clustering of large volatility events in financial time series data\citep{cont2001, Harras2012}. Ising like models of opinion dynamics are often used to model returns time series and are known to be able to reproduce intermittent dynamics and volatility clustering observed in returns data\citep{Bornholdt2001, Giardina2003, Cont2005}. In this work, we are particularly interested in a mean-field version of the Ising model which provides a concise, discrete-time model of opinion dynamics in financial markets. One of the mechanisms underlying volatility clustering in opinion dynamics has been demonstrated to be similar to attractor bubbling in dynamical systems theory\citep{Krawiecki2002}. Attractor bubbling is known to occur in dynamical systems possessing a lower dimensional invariant manifold with subsets that are unstable\citep{Ashwin1994,Ott1996}. When trajectories of such systems pass through a region of instability, any perturbation away from the invariant manifold grows and causes the trajectory to deviate from its original path on the invariant manifold\cite{Cavalcante2013, Cartwright2002}. Depending on the extent of such unstable sets, a system may exhibit highly intermittent behavior which resembles volatility clustering commonly seen in financial time series data.

To investigate the phenomenon of volatility clustering in mean-field Ising models of opinion dynamics, we use a bailout embedding\citep{Cartwright2002}. The bailout embedding is an analytical technique inspired by the dynamics of finite sized particles in fluids\citep{MR1983,Cartwright2010} and was developed to enable the study of dynamical systems with an underlying structure that gives rise to behaviors such as attractor bubbling and intermittency\citep{Cartwright2002, Cartwright2003}. Using a bailout embedded version of the discrete-time map under study, we are able to classify subsets of phase space as either stable or unstable to perturbations for a given range of parameters. This allows us to understand the role played by these phase space regions in determining the overall dynamics of the system. 
At this point it is important to emphasize that the connection between attractor bubbling and volatility clustering has been known for over two decades\cite{Krawiecki2002} and this work investigates this connection further for a particular model of interest within the framework of a bailout embedding. Doing this exercise offers us novel insights into the relationship between an inertia like parameter and the onset of volatility clustering in collective dynamics. In addition, some observations are made about the nature of forcing required to induce intermittent dynamics.
\newline
The paper is organized as follows. Section \ref{section:bom} presents the discrete-time mean-field Ising model chosen for study in this paper. This section also introduces the idea of a bailout embedding, followed by its application to the discrete time map under consideration. Section \ref{section:stability} demonstrates the application of transversal Lypaunov exponents to classify the underlying phase space into stable and unstable sets. The influence of the stable and unstable sets on the dynamics of individual trajectories is presented for the case of an isolated perturbation \big(subsection (A)\big) as well as the case of sustained perturbation \big(subsection (B)\big). We also demonstrate the role of the bailout function in the onset of volatility clustering. In Section \ref{section:bof}, we draw parallels between the dynamics of the bailout embedded Ising model and the dynamics of inertial particles in fluids. This leads us to an interpretation of the bailout function as a parameter that quantifies investor inertia.  Finally, in Section \ref{section:conclusion} we offer some concluding thoughts while the Appendix shows details of bailout embedding of the system in (\ref{eq:1}).
%
%

\section{Description and bailout embedding of a 2D map}\label{section:bom}

\begin{figure*}
\centering
\begin{minipage}{0.5\linewidth}
    \centering
    \includegraphics[width=\linewidth]{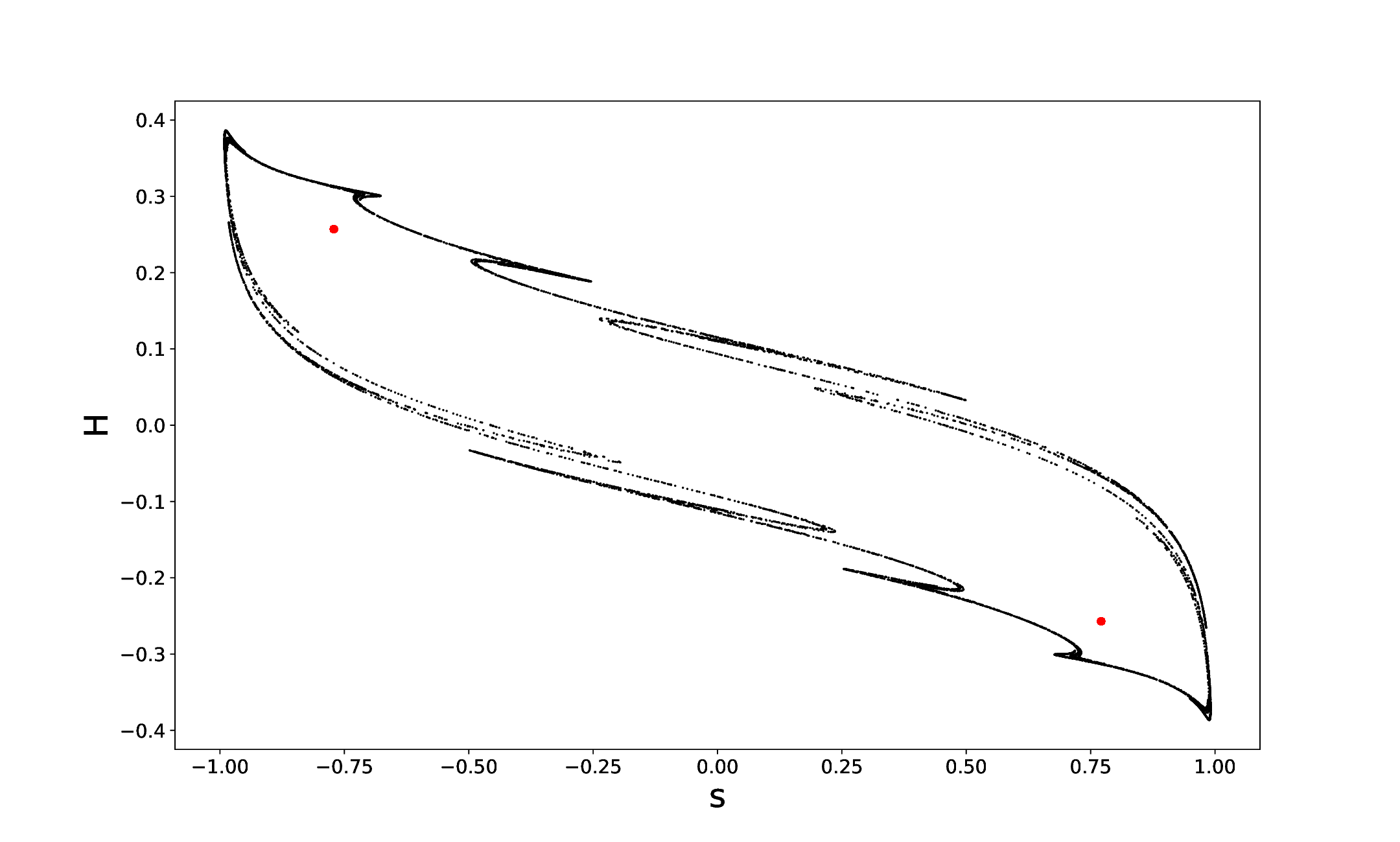}

\end{minipage}%
\hspace{0.05\linewidth} 
\begin{minipage}{0.7\linewidth}
    \centering
    \includegraphics[width=\linewidth]{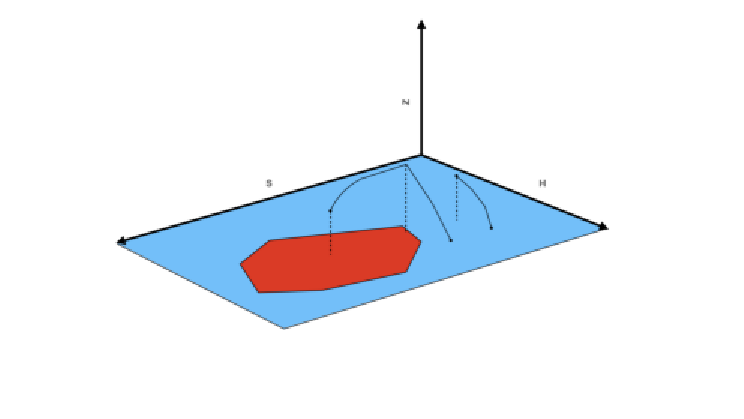}
 
\end{minipage}
\caption{(a) Chaotic attractor(black) co-exists with a periodic attractor(red).(b) Schematic: ($\sigma$) quantifies the rate of growth of perturbation $Z=\sqrt{u^2+v^2}$ at any point $\big(s,H\big)$, with $\sigma \big(s,H\big)>0$ on unstable points and $\sigma \big(s,H\big)<0$ on stable points on the invariant manifold.}
\label{fig:atttractor}
\end{figure*}

We begin with the description of a 2D map that was originally presented in \cite{Smug2018}. The map is given by,
\begin{eqnarray}\label{eq:1}
    s(n+1)=\tanh \big( a\cdot s(n)+b\cdot H(n)\big)\\
    H(n+1)= \theta\cdot H(n)+(1-\theta)s(n)\nonumber
\end{eqnarray}
where $s(n)$ is the state variable reflecting the prevailing opinion at iteration $n$, of a representative agent in the Ising model, with positive values denoting buy and negative values denoting sell. The variable $H(n)$ denotes the influence of the market's momentum or trend on the agent's behavior and is computed as an exponential moving average of the previous $\big(\frac{1}{1-\theta}\big)$ time steps in the opinion states. Thus the parameter $\theta \in [0,1)$ is a measure of memory in the variable $H(n)$ such that for $\theta=0$, one gets $H(n+1)=s(n)$ (one-step memory) and as $\theta\rightarrow{1^{-}}$ we get $H(n+1)=H(n)=H(0)$ (infinite memory). The constant $a$ quantifies the feedback effect or the inclination of an agent to imitate another, on the dynamics, and $b$ quantifies an organizing effect on the dynamics due to the field $H$. The authors of \citep{Smug2018} have demonstrated that this system exhibits rich dynamics in phase space, involving several possible stable and unstable equilibria, bifurcations and chaos. In this work we confine ourselves to one set of parameters $a=-4.17,b=-8.53, \theta=0.5$ for which the system is known to simultaneously possess a chaotic and a periodic attractor\citep{Smug2018} seen in Fig.\ref{fig:atttractor}(a). 

The authors of \citep{Cartwright2003} address the question of stability to perturbation of attractors embedded within lower-dimensional invariant manifolds in phase space. They use bailout embeddings as a tool to arrive at thresholds for parameter values above which either some subsets of the attractor or the entire attractor can become unstable leading to bubbling and blowout bifurcations respectively, of the attractor. 

The bailout embedding\cite{Cartwright2002} of a map $x_{n+1}=f(x_n)$ is a map of the form $x_{n+2}-f(x_{n+1})=K(x_n)[x_{n+1}-f(x_n)]$ where $K(x_n)=e^{-\gamma}\frac{\partial f}{\partial x_n}$ is called the bailout function and $\gamma$ is a bailout parameter. Applying the idea of bailout embedding to the 2D system of maps in (\ref{eq:1}), we obtain the following 4-D system(See Appendix for detailed calculations),

\begin{align}\label{eq:2}
    & s(n+1)=u(n)+\tanh \big(a\cdot s(n)+b\cdot H(n)\big)\\
    & H(n+1)= v(n)+\theta\cdot H(n)+(1-\theta)s(n)\nonumber\\
    & u(n+1)=e^{-\gamma}[a\sech^2\big(a\cdot s(n)+b\cdot H(n)\big)u(n) \ldots\nonumber\\
    & +b\sech^2\big(a\cdot s(n)+b\cdot H(n)\big)v(n)]\nonumber\\
    & v(n+1)=e^{-\gamma}[(1-\theta)u(n)+\theta v(n)]\nonumber.
\end{align}

The plane $\big(s,H,u=0,v=0\big)$ is thus an invariant manifold of the full 4-D dynamics in the $\big(s,H,u,v\big)$ space, and represents the unperturbed opinion dynamics. It was shown in \cite{Krawiecki2002}, that a 1D mean-field Ising model of opinion dynamics exhibits attractor bubbling, which leads to discrete opinion dynamics resembling volatility clustering when the model is perturbed by uniformly distributed noise. In what follows, we study the stability of the $4-D$ system (\ref{eq:2}) by studying its response to an isolated perturbation followed by a study of the dynamics in the presence of sustained, noise induced perturbation.
\begin{figure*}
\centering
\begin{minipage}{0.45\linewidth}
    \centering
    \includegraphics[width=\linewidth]{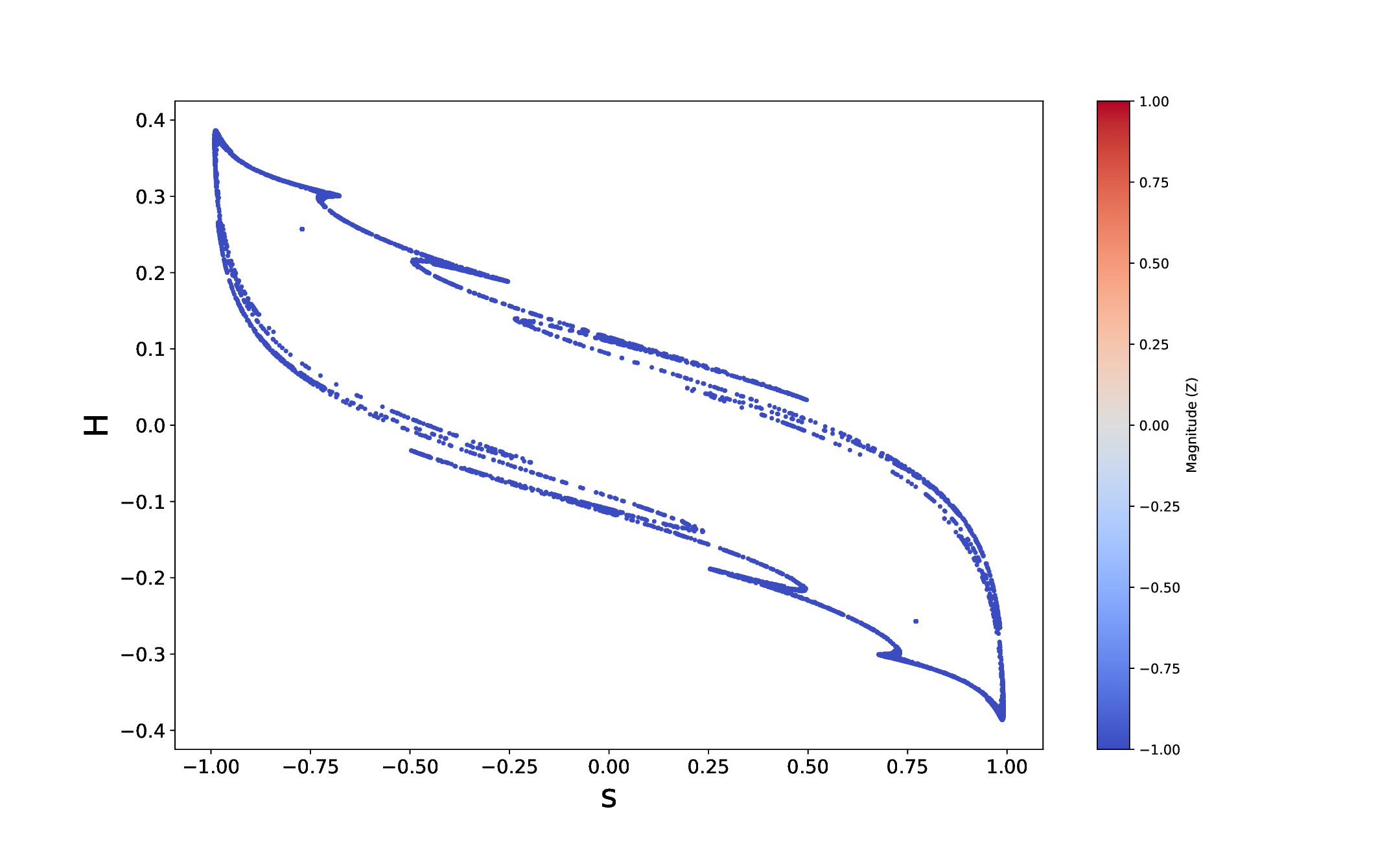}
    (a) $\gamma=10$
\end{minipage}%
\hspace{0.05\linewidth} 
\begin{minipage}{0.45\linewidth}
    \centering
    \includegraphics[width=\linewidth]{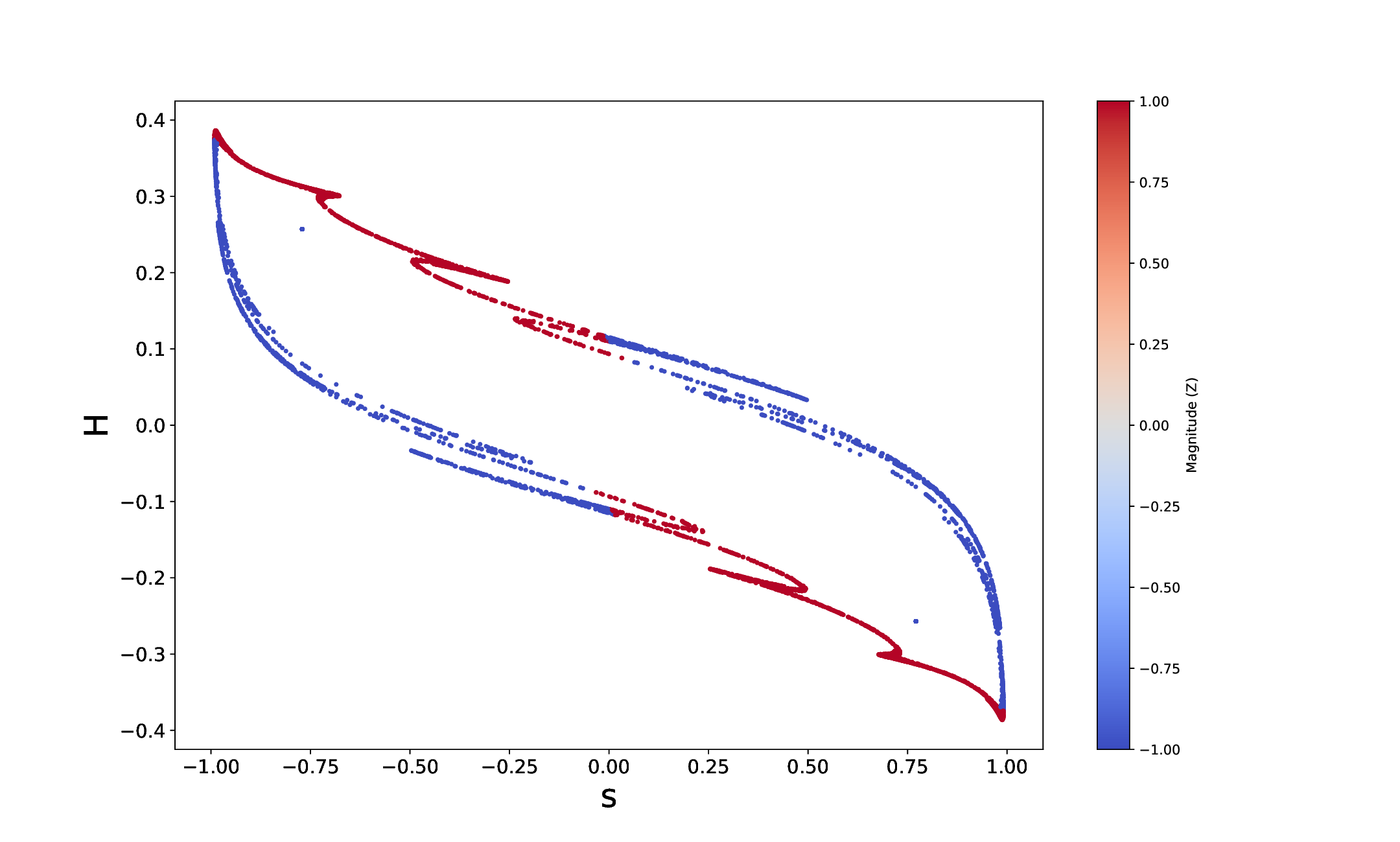}
    (b) $\gamma=1$
\end{minipage}

\vspace{0.05\linewidth} 

\begin{minipage}{0.45\linewidth}
    \centering
    \includegraphics[width=\linewidth]{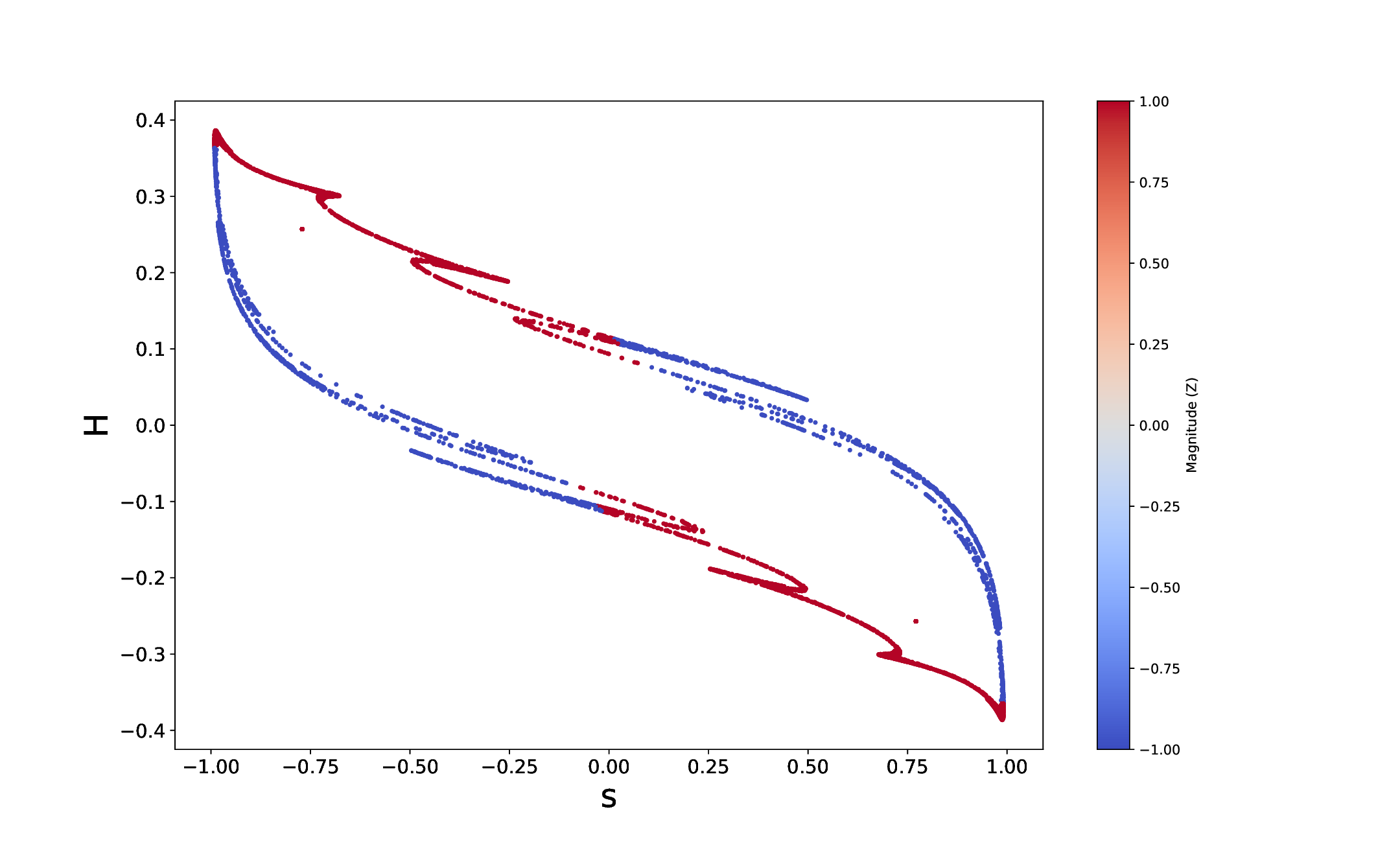}
    (c) $\gamma=0.83$
\end{minipage}%
\hspace{0.05\linewidth} 
\begin{minipage}{0.45\linewidth}
    \centering
    \includegraphics[width=\linewidth]{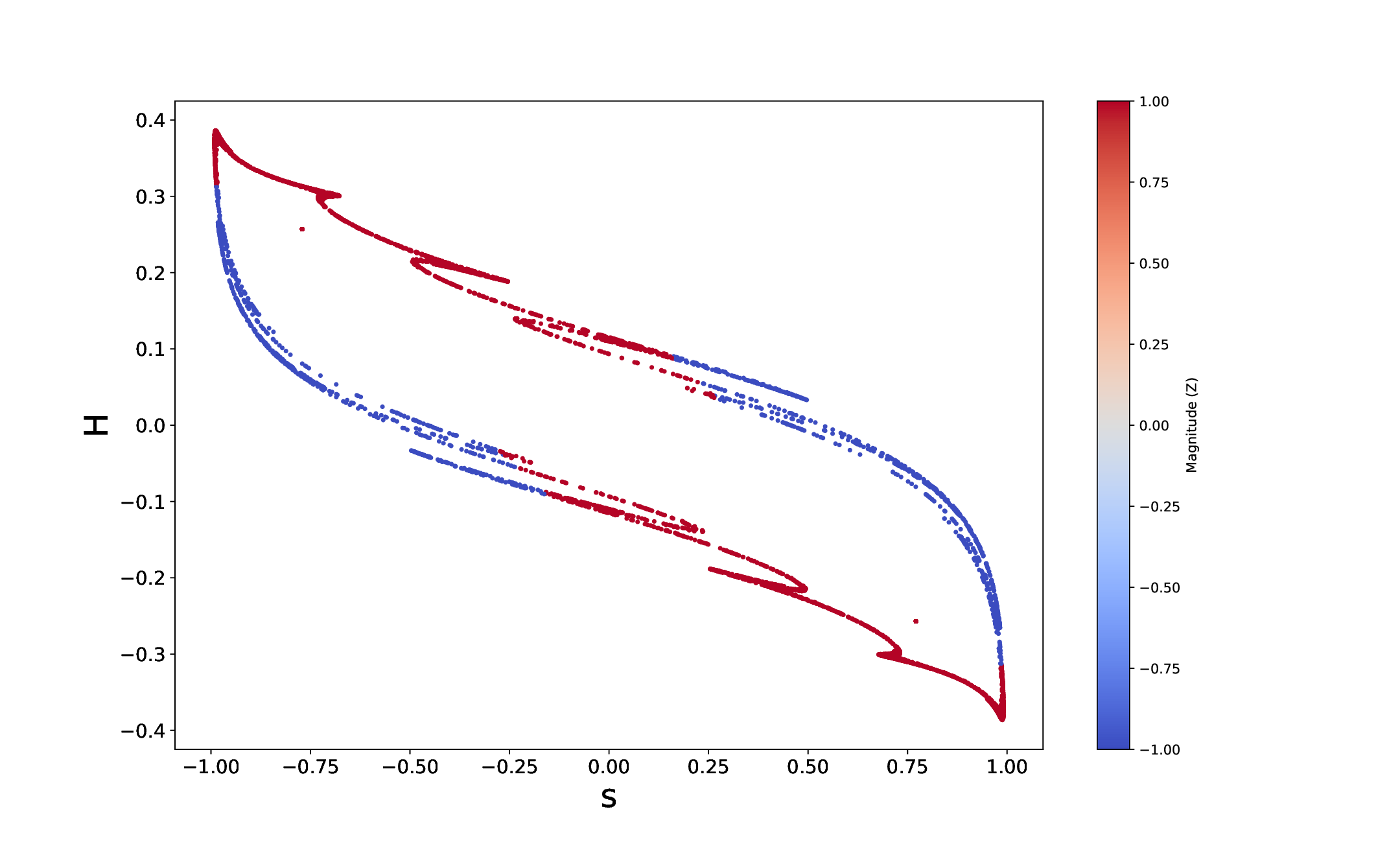}
    (d) $\gamma=0.1$
\end{minipage}
\caption{Stable points are indicated in blue and unstable points are in red. (a) The chaotic and periodic attractors are stable. (b) The periodic attractor is stable and a subset of the chaotic attractor loses stability as $e^{-\gamma}$ is increased. (c) The periodic attractor loses stability. (d) The extent of the unstable set in phase space for a large value of the bailout function.}
\label{fig:nile_plots}
\end{figure*}

\section{Stability analysis of the embedded map}\label{section:stability}
Invariant manifolds in the phase space of certain dynamical systems may contain riddled basins of attraction\citep{Alexander1992, Ashwin1996} leading to dynamics such as intermittency\citep{Ott1994} and blow-out bifurcations\citep{Ott1994_1}. To measure the instability associated with an orbit on the invariant manifold, the authors of \citep{Cartwright2003} use a transversal Lyapunov exponent, which is an asymptotic quantity that is positive on average on unstable orbits and negative on average on stable orbits as $n \rightarrow{\infty}$ ( with $n$ being the number of iterations of the discrete time dynamical system). In this work we are concerned with transient instabilities associated with chaotic and periodic orbits that cause perturbed trajectories to be temporarily repelled from an orbit at one point but may still settle down onto another point on the same orbit eventually. This can occur after the associated perturbation of magnitude $Z=\sqrt{u(n)^2+v(n)^2}$ dissipates completely upon the trajectory visiting a stable portion of the same orbit. To identify such points of transient instability normal to the invariant manifold, we compute a modified version of the transversal Lyapunov exponent which can be understood as a discrete time version of the normal infinitesimal Lyapunov exponent\cite{Haller2010} and is defined as,
\begin{eqnarray}\label{eq:3}
    \sigma(s,H)= \log[\frac{\delta_1}{\delta_0}]
\end{eqnarray}
where $\delta_0=\sqrt{u(0)^2+v(0)^2}$ is the magnitude of the initial perturbation in the ($u-v$) space and $\delta_1=\sqrt{u(1)^2+v(1)^2}$ is the magnitude of the perturbation after one iteration of the map (\ref{eq:2}). Any perturbation normal to the invariant manifold thus grows as long as it is associated with an unstable point$(s^u,H^u)$ such that $\sigma(s^u,H^u)>0$ and decays at stable points$(s^s,H^s)$ with $\sigma(s^s,H^s))<0$ on the invariant manifold, as depicted in the schematic in Fig.\ref{fig:atttractor}(b). Using this criteria, one can partition the chaotic and periodic attractors on the invariant manifold into an unstable subset with $\sigma(s,H)>0$ at every point within the set and a stable subset with $\sigma(s,H)<0$ on every point within the set. To do this, we first evolve a rectangular grid of points in the $(s-H)$ state space for an arbitrarily long time such that the dynamics converges onto the chaotic and periodic orbits from within their respective basins of attraction. We then perform a normal stability computation at these attracting points using the measure described above in (\ref{eq:3}) followed by thresholding the values of $\sigma$ with points $(s, H)$ such that $\sigma(s, H)>0$ being assigned a value of $+1$ and a value of $-1$ otherwise. This reveals the stable and unstable points as different ends of the color spectrum on the color map. In what follows, we will study the variation of the extent of such stable and unstable sets as the bailout function ($e^{-\gamma}$) is varied.

We start our numerical experiment for small values of the bailout function $e^{-\gamma}$(large $\gamma$). We compute the discrete-time, infinitesimal normal Lyapunov exponent over one iteration starting at every point on the chaotic and periodic attractors and use its value to identify stable and unstable points as described in the preceding paragraph. For values of $\gamma\approx 10$, we see that both, the chaotic and periodic attractors are completely stable as indicated using blue bubbles at the respective locations on the attractors Fig. \ref{fig:nile_plots}(a). Upon increasing the value of the bailout function gradually, the chaotic and periodic attractors remain stable until $\gamma\approx 2.1$, at which point the chaotic attractor loses stability at a few points, while other points on the chaotic attractor as well as the periodic attractor retain stability. As $e^{-\gamma}$ is further increased, the extent of the unstable set on the chaotic attractor increases slightly while the periodic attractor retains stability. A representative plot of this state($\gamma<2.1$) is shown in Fig. \ref{fig:nile_plots}(b). The next bifurcation occurs at $\gamma\approx 0.83$ when the periodic attractor loses stability and becomes unstable as seen in Fig. \ref{fig:nile_plots}(c) (red bubbles). Further increasing the value of $e^{-\gamma}$ increases the extent of the unstable set on the chaotic attractor, but it does not qualitatively change the picture of stability on the invariant manifold as demonstrated for $\gamma=0.1$ in Fig. \ref{fig:nile_plots}(d). 
\subsection{Dynamics and stability for an isolated perturbation}
Having understood the global picture of stability and bifurcations that occur in the system as the bailout function is systematically varied, we turn our attention to the dynamics of individual trajectories. Studying the behavior of individual trajectories in the backdrop of stable and unstable sets computed previously helps us understand implications of the bifurcations on the system's dynamics, in addition to highlighting any anomalous behavior that is not easily explained using the stability analysis conducted previously. To be able to study stability and convergence to the chaotic and periodic attractors separately, it is useful to identify the boundaries of their respective basins of attraction. We do this by computing the finite time Lyapunov exponent(FTLE) field for dynamics on the invariant manifold $(s,H,0,0)$ of the system in (\ref{eq:2}) which is equivalent to studying the dynamics of system in (\ref{eq:1}). The FTLE is a method to identify transport barriers in phase space that has found application in the study of numerous problems in science and engineering\citep{Haller2015}. In what follows, we review the definition of FTLE following the exposition in the works of Shadden and co-workers \citep{Shadden2005,Shadden_LCS}.

Consider a map $\phi_{t_0}^{t_0+T}(x): \mathbb{R}^n\rightarrow{\mathbb{R}^n}$ and two points in phase space separated by a distance of $\delta x(t_0)$, mapped forward by this map. After a time $T$ the separation between the points is given as,

\begin{multline*}
    \delta x(t_0+T) = \phi_{t_0}^{t_0+T}\big(x+\delta x(t_0)\big)-\phi_{t_0}^{t_0+T}(x)\\
    =\frac{d\phi_{t_0}^{t_0+T}(x)}{dx}\delta x(t_0)+\mathcal{O}(\| \delta x(t_0) \|^2).
\end{multline*}
The magnitude of growth of the perturbation is then given as,
\begin{multline*}
    \|\delta x(t_0+T)\|=\sqrt{<\delta x(t_0),\frac{d\phi_{t_0}^{t_0+T}(x)}{dx}^*\frac{d\phi_{t_0}^{t_0+T}(x)}{dx}\delta x(t_0)>}
\end{multline*}
with the superscript $^*$ denoting transpose of the operator. Here the matrix $C=\frac{d\phi_{t_0}^{t_0+T}(x)}{dx}^*\frac{d\phi_{t_0}^{t_0+T}(x)}{dx}$ is the (finite-time) Cauchy-Green strain tensor. The leading eigenvector of $C$ gives the direction of maximal stretching in phase space, and the magnitude of maximal stretching is then given by,
\begin{eqnarray*}
     max\|\delta x(t_0+T)\|=\sqrt{<\overline\delta x(t_0),(\lambda_{max}C)\overline\delta x(t_0)>}\\
     =\sqrt{\lambda_{max}C}\|(\overline\delta x(t_0)) \|
\end{eqnarray*}
where $\lambda_{max}C$ is the maximum eigenvalue of $C$ and $\overline\delta x(t_0)$ is aligned with the associated  eigenvector.

The FTLE over a finite time interval $T$ is then defined as,
\begin{equation}\label{eq:4}
    (\sigma_f(x))_{t_0}^{T}=\frac{1}{|T|}\ln(\sqrt{\lambda_{max}C})
\end{equation}
so that the maximal of growth of $\delta x(t_0)$ can now be written as,

\begin{equation*}
     max\|\delta x(t_0+T)\|=e^{\big(\sigma_f(x)_{t_0}^{T}\big)|T|}\| \overline\delta x(t_0) \|.
\end{equation*}

\begin{figure}
\centering
    \includegraphics[width=\linewidth]{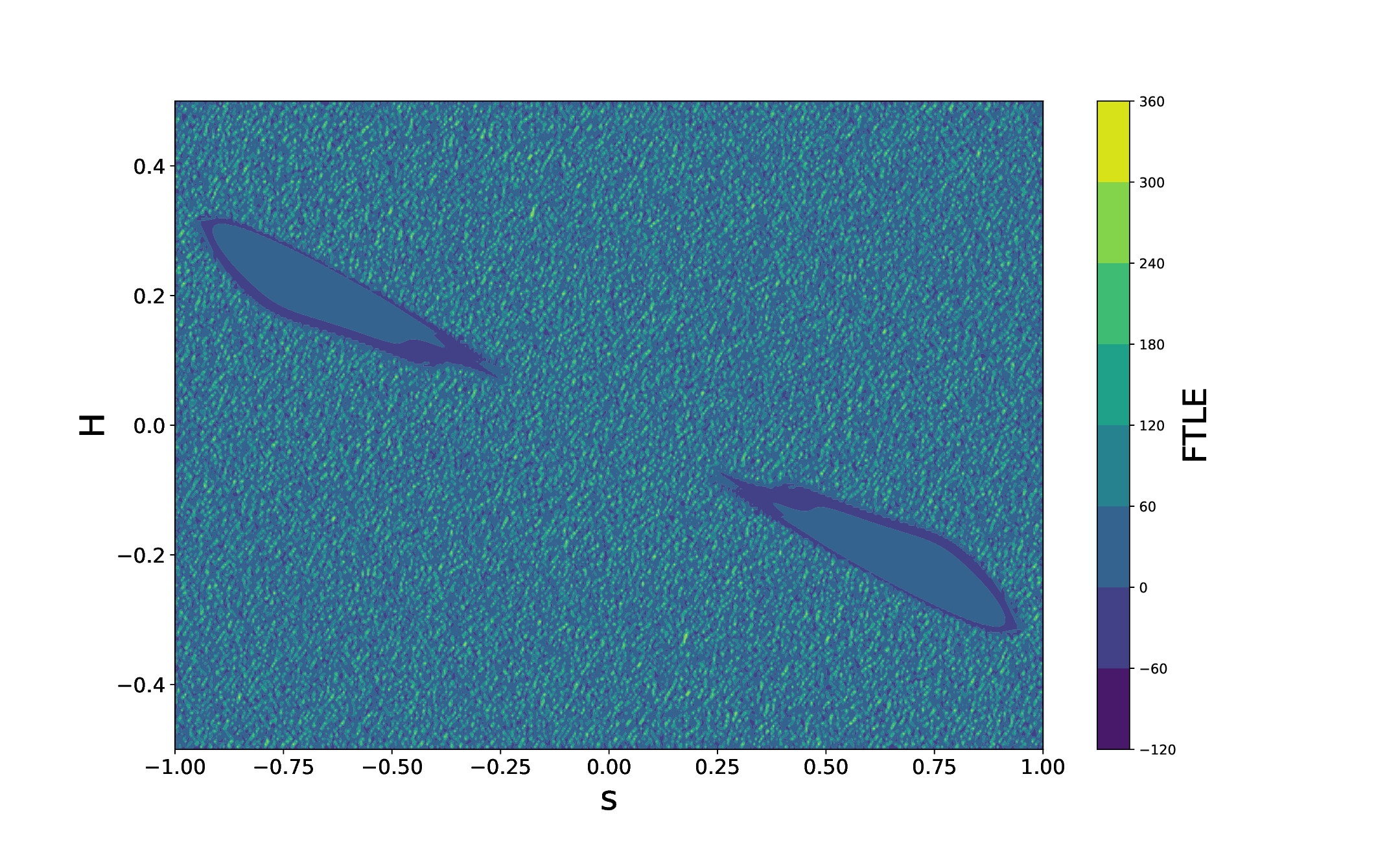}
\caption{The finite time Lyapunov exponent(FTLE) field reveals regions of qualitatively different dynamics.}
\label{fig:FTLE}
\end{figure}

With FTLE having been defined in (\ref{eq:4}), we can now compute the FTLE field for the map over a finite time interval as seen in Fig.\ref{fig:FTLE}, where the finite time interval is taken to be 100 iterations of the map. One can now recognize a region of relatively low values of scalar FTLE surrounding the periodic attractor and a second region with higher values of the FTLE surrounding the first region. The high FTLE ridge acts as a barrier to transport in phase space and thereby identifies a separatrix that separates periodic and chaotic dynamics. It is worth emphasizing at this point that the value of $\gamma$ does not affect the FTLE field qualitatively owing to the fact that the FTLE computation is performed for the zero disturbance($Z=0$) case, however, it provides us with a starting point for our study of perturbed dynamics with $Z\neq 0$.

\begin{figure*}
\centering
\begin{minipage}{0.4\linewidth}
    \centering
    \includegraphics[width=\linewidth]{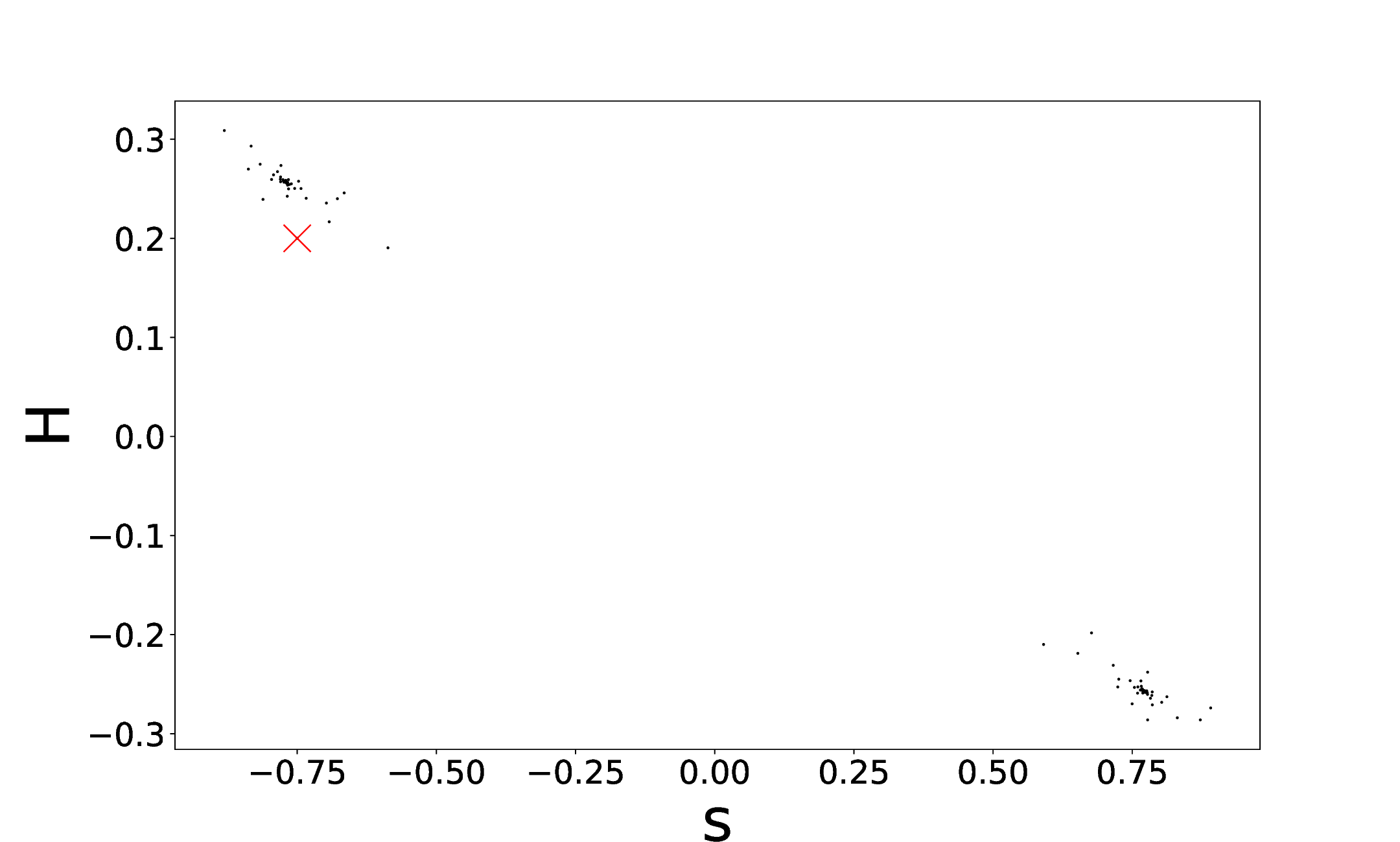}
\end{minipage}%
\hspace{0.05\linewidth} 
\begin{minipage}{0.4\linewidth}
    \centering
    \includegraphics[width=\linewidth]{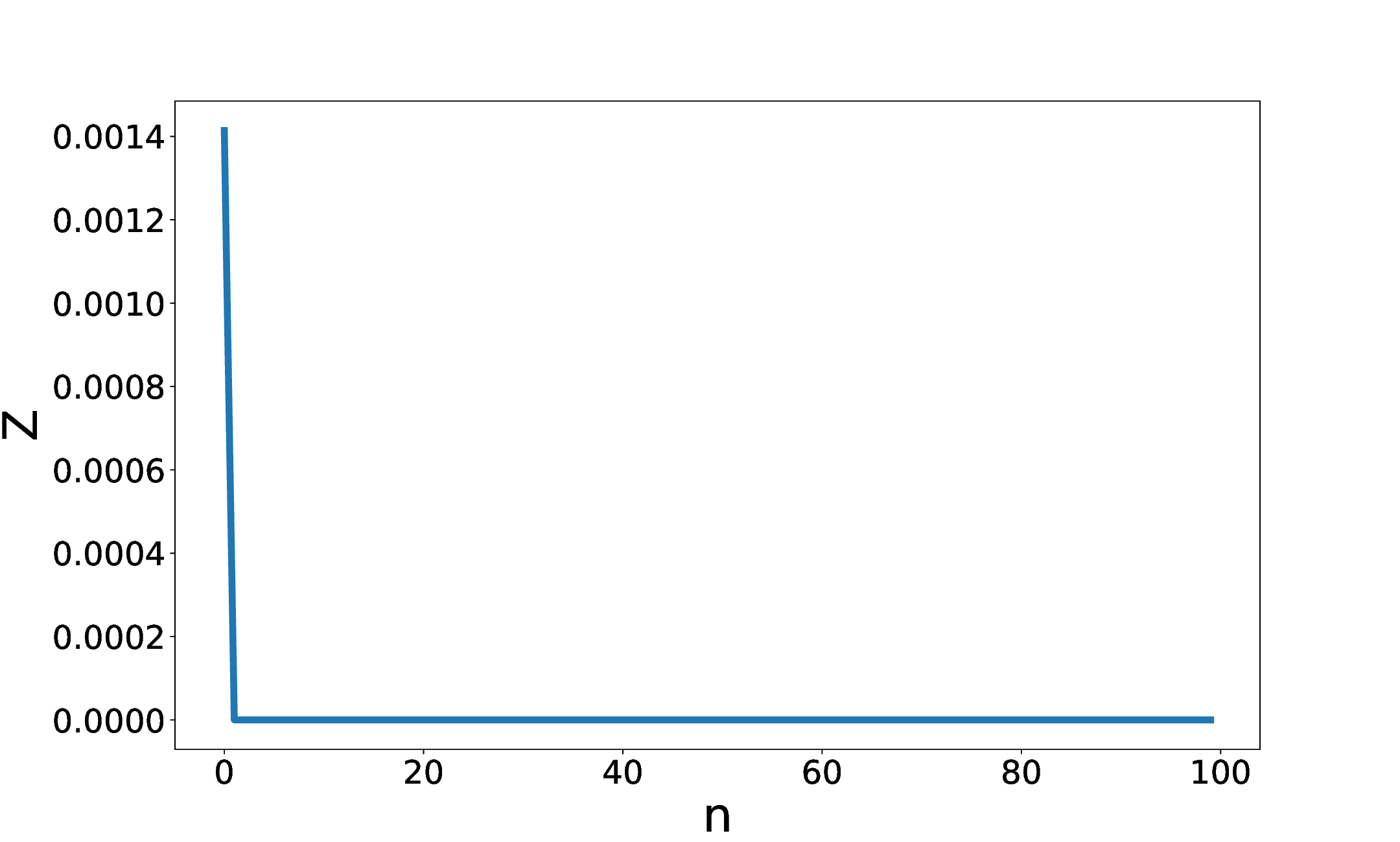}
\end{minipage}

\centering 
(a) $\gamma=10$ with $(s, H)=(-0.75,0.2)$
\\
\begin{minipage}{0.4\linewidth}
    \centering
    \includegraphics[width=\linewidth]{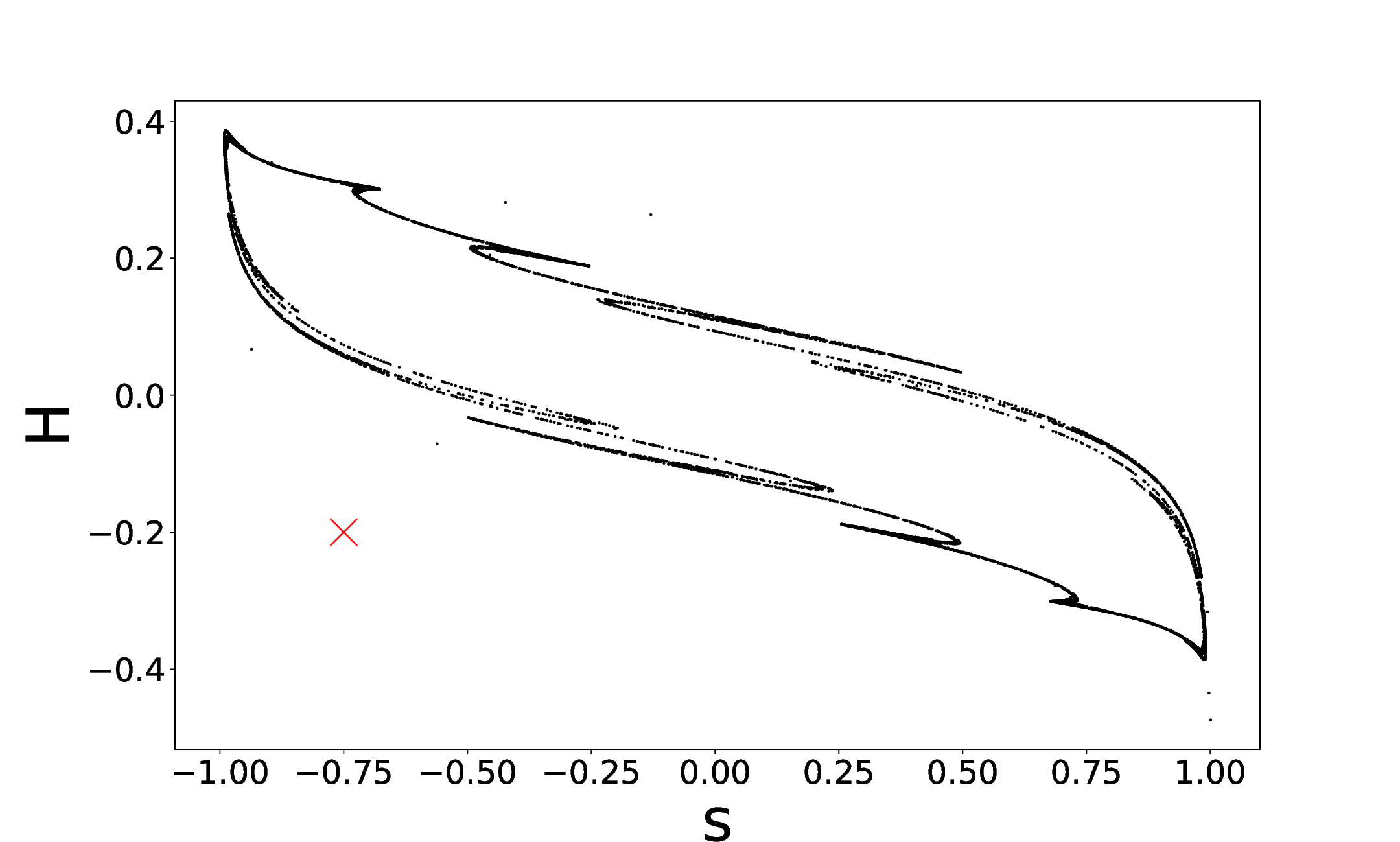}
\end{minipage}
\hspace{0.05\linewidth} 
\begin{minipage}{0.4\linewidth}
    \centering
    \includegraphics[width=\linewidth]{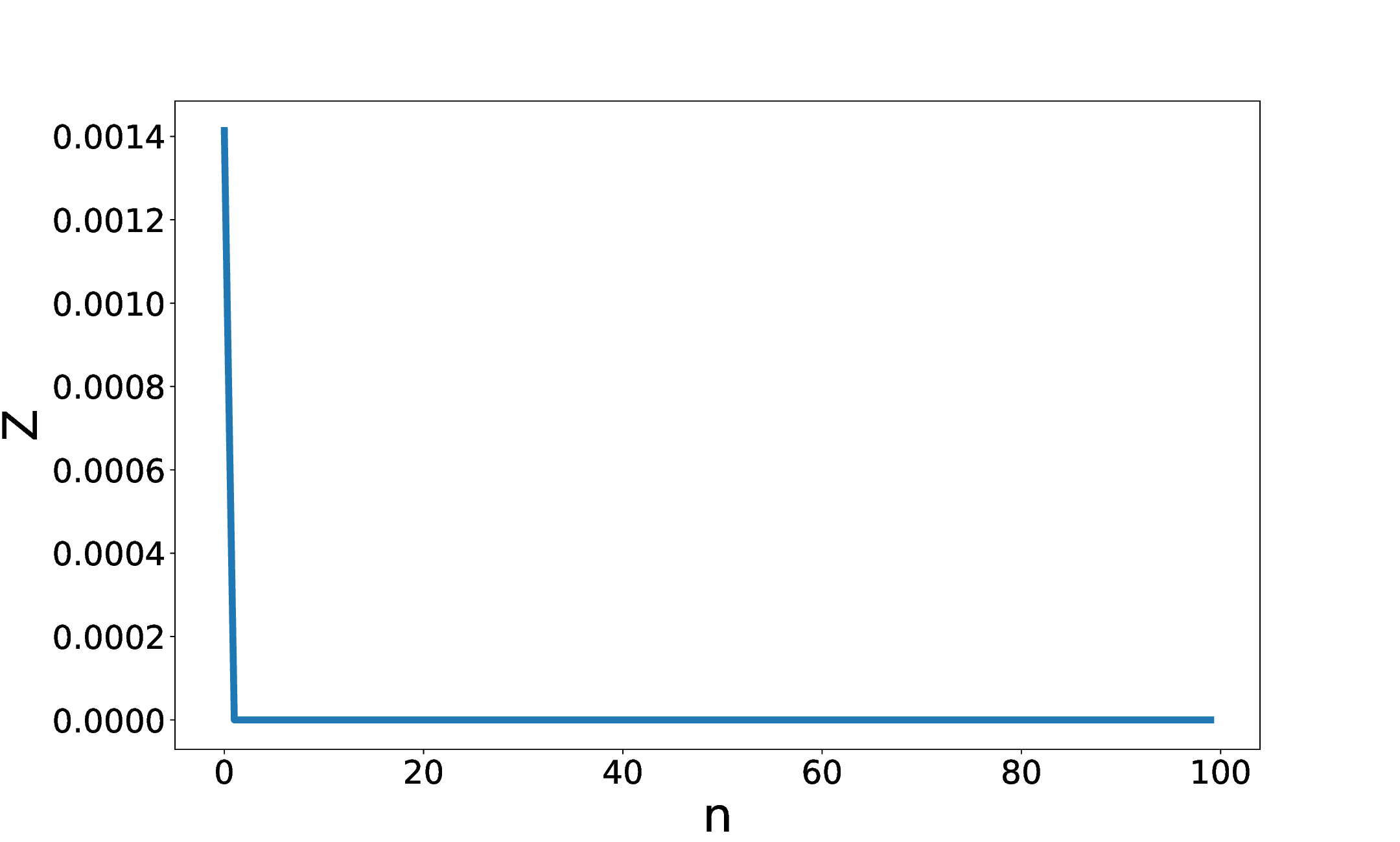}
\end{minipage}%

\centering 
(b) $\gamma=10$ with $(s, H)=(-0.75,-0.2)$
\\
\begin{minipage}{0.4\linewidth}
    \centering
    \includegraphics[width=\linewidth]{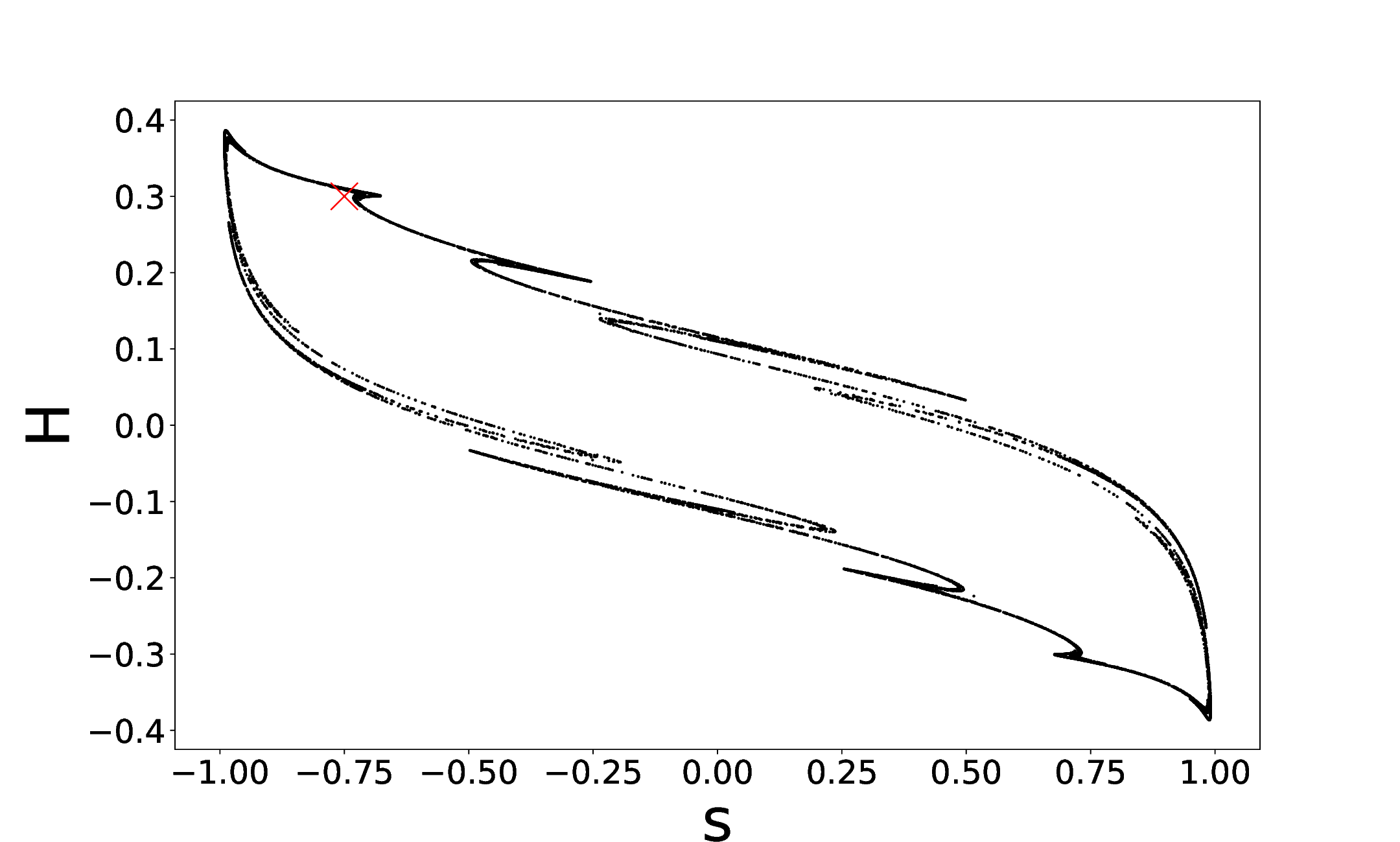}
\end{minipage}
\hspace{0.05\linewidth} 
\begin{minipage}{0.4\linewidth}
    \centering
    \includegraphics[width=\linewidth]{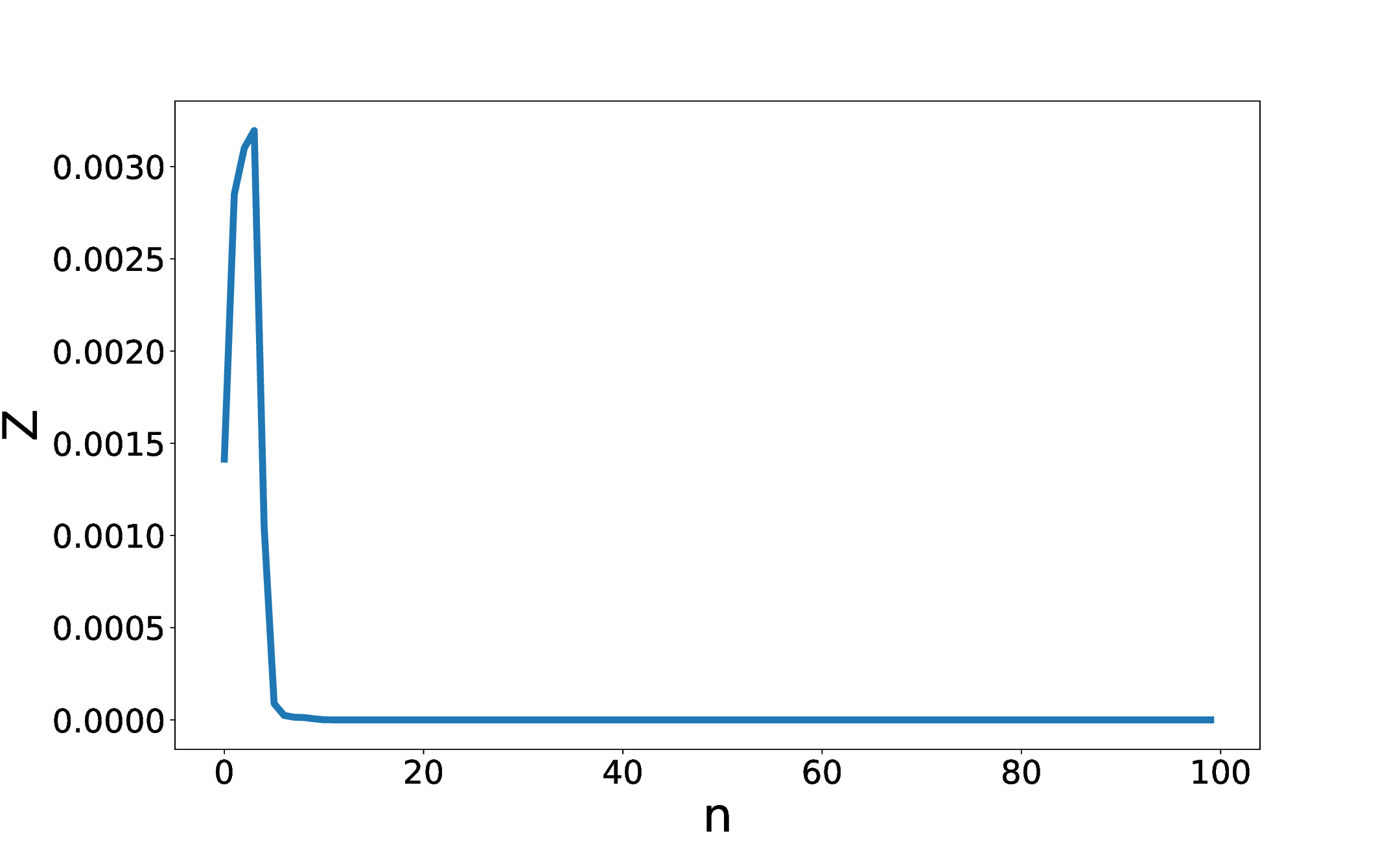}
\end{minipage}

\centering 
(c) $\gamma=1$ with $(s, H)=(-0.75,0.3)$
\\
\begin{minipage}{0.4\linewidth}
    \centering
    \includegraphics[width=\linewidth]{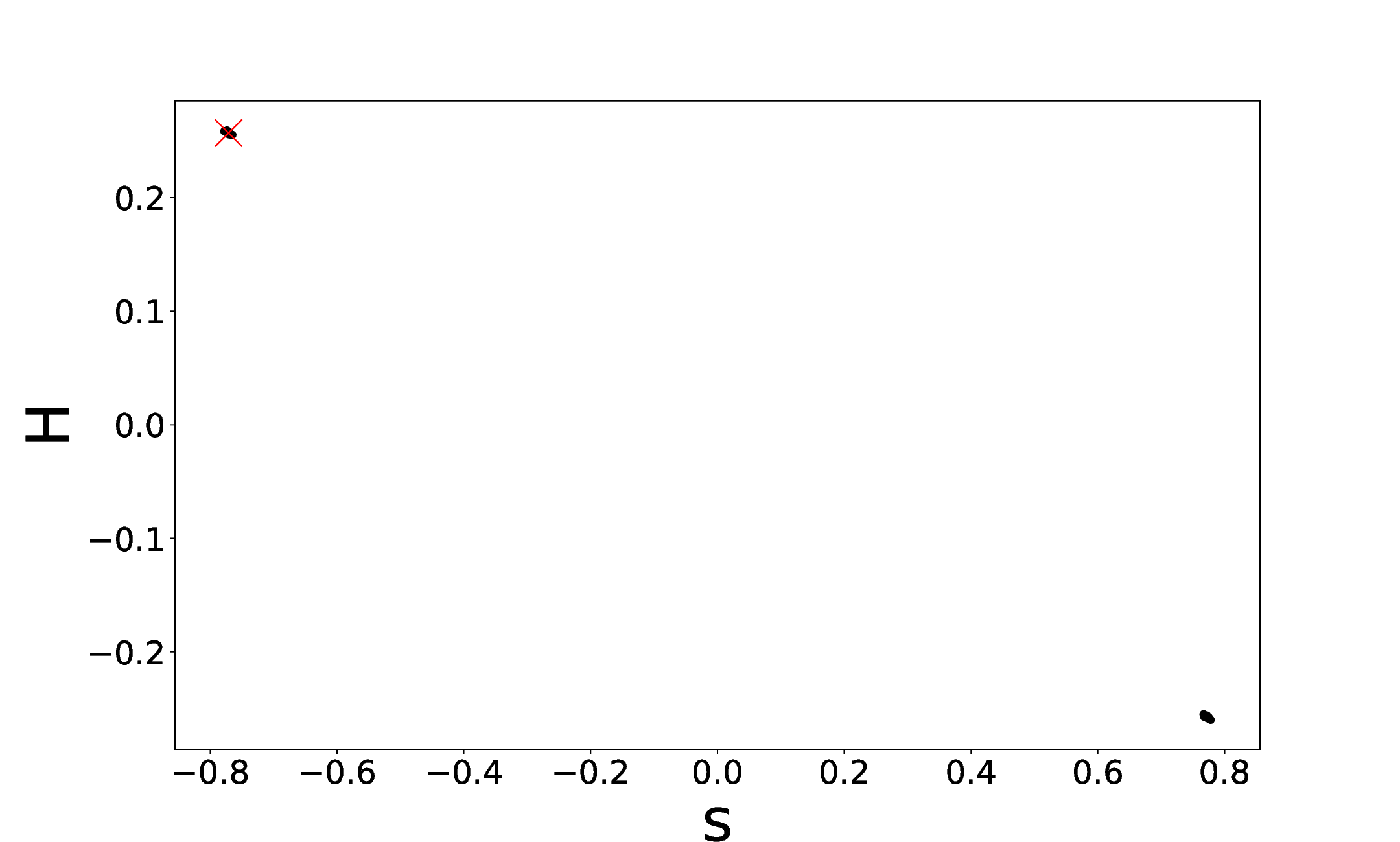}
\end{minipage}%
\hspace{0.05\linewidth} 
\begin{minipage}{0.4\linewidth}
    \centering
    \includegraphics[width=\linewidth]{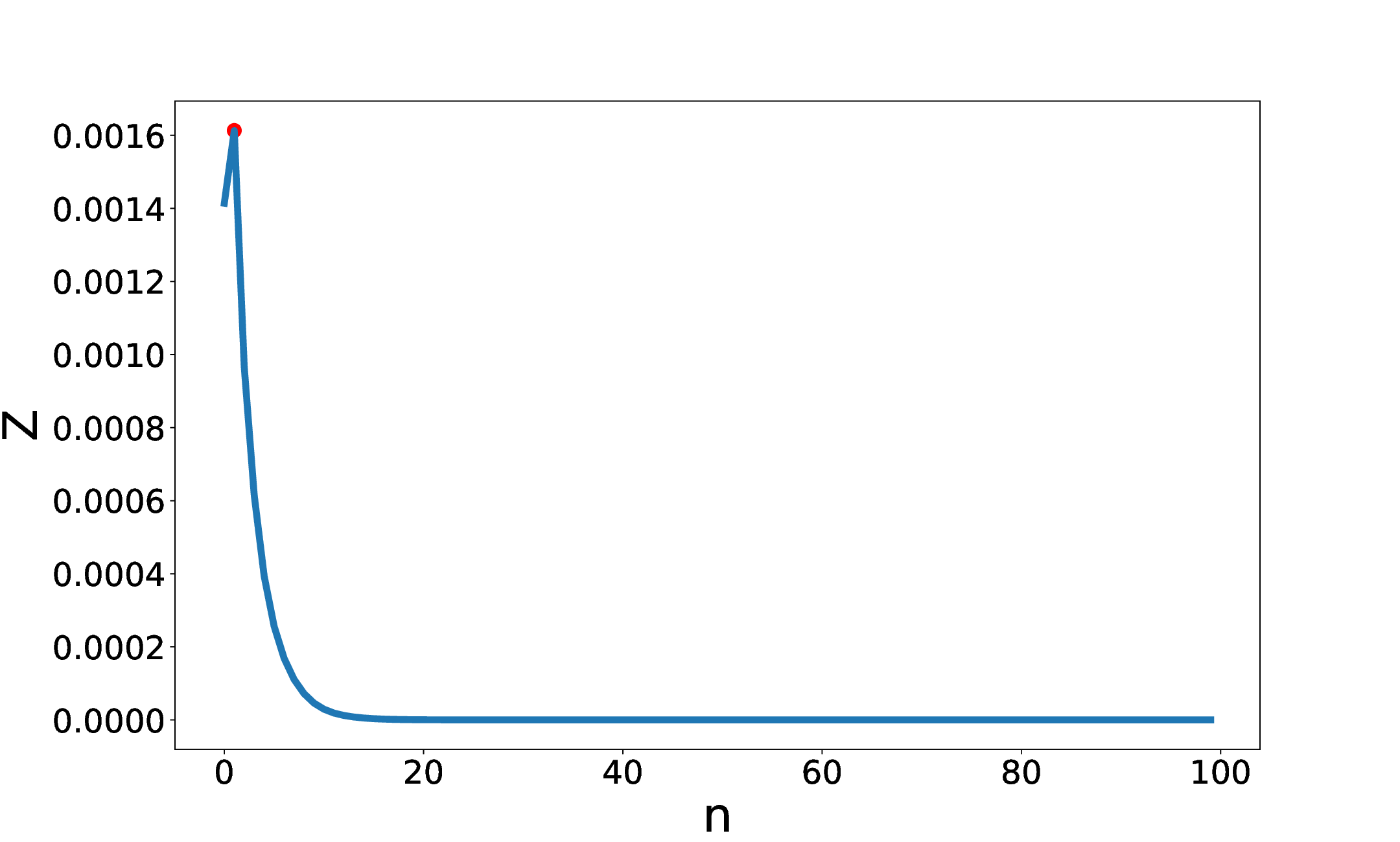}
\end{minipage}%

\centering 
(d) $\gamma=0.7$ with $(s, H)=(-0.7711,0.2570)$
\caption{(Left) Convergence of the map to either the periodic or the chaotic attractor depending on the initial condition. The location of the initial condition supplied to map (\ref{eq:2}) is marked with a red cross for each case. (Right) The decay of perturbation $Z=\sqrt{u^2+v^2}$ against iteration number $n$, for each case.}\label{fig:traj}
\end{figure*}

\begin{figure*}
\begin{minipage}{0.4\linewidth}
    \centering
    \includegraphics[width=\linewidth]{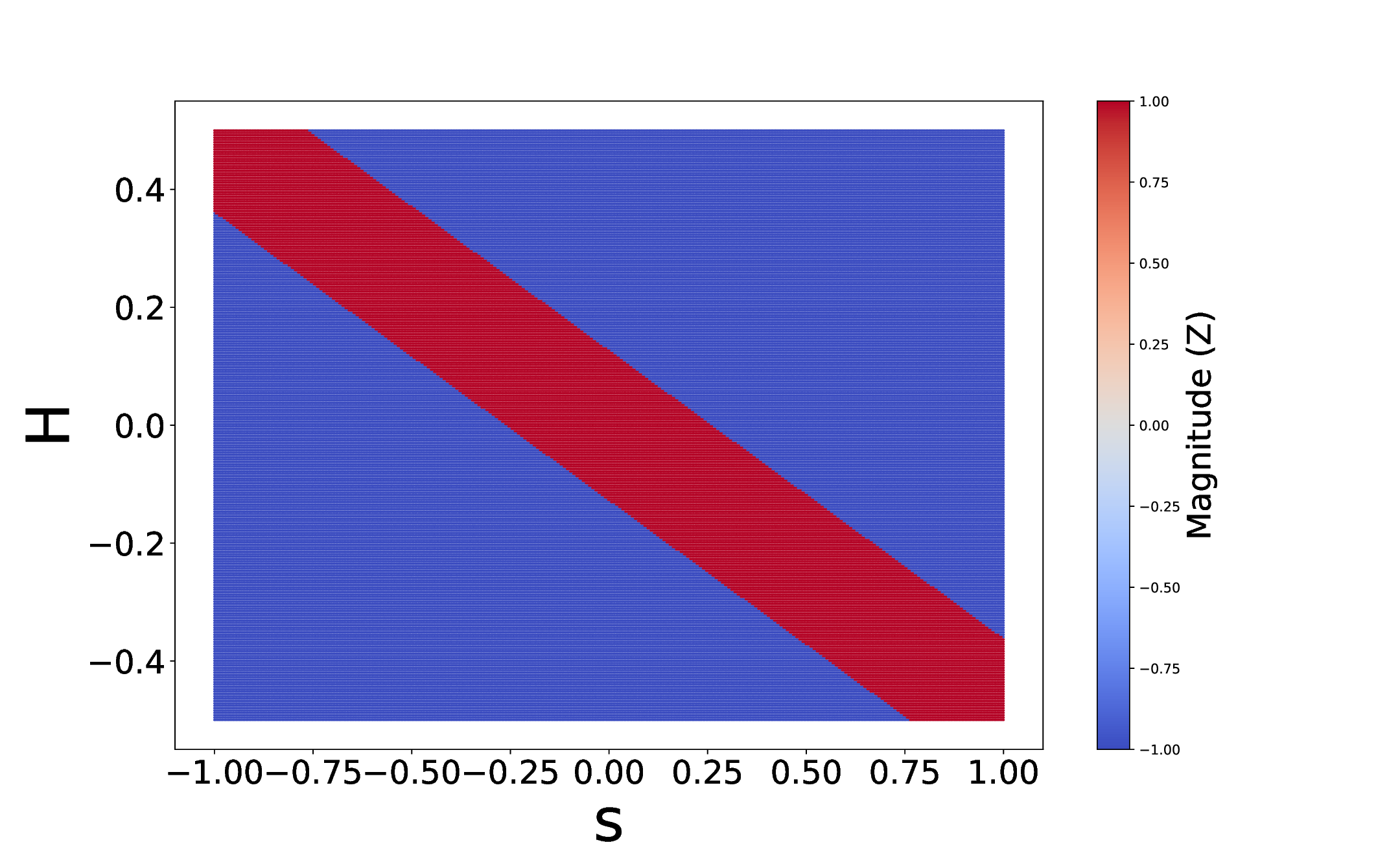}
    (a)
\end{minipage}
\hspace{0.1\linewidth} 
\begin{minipage}{0.4\linewidth}
    \centering
    \includegraphics[width=\linewidth]{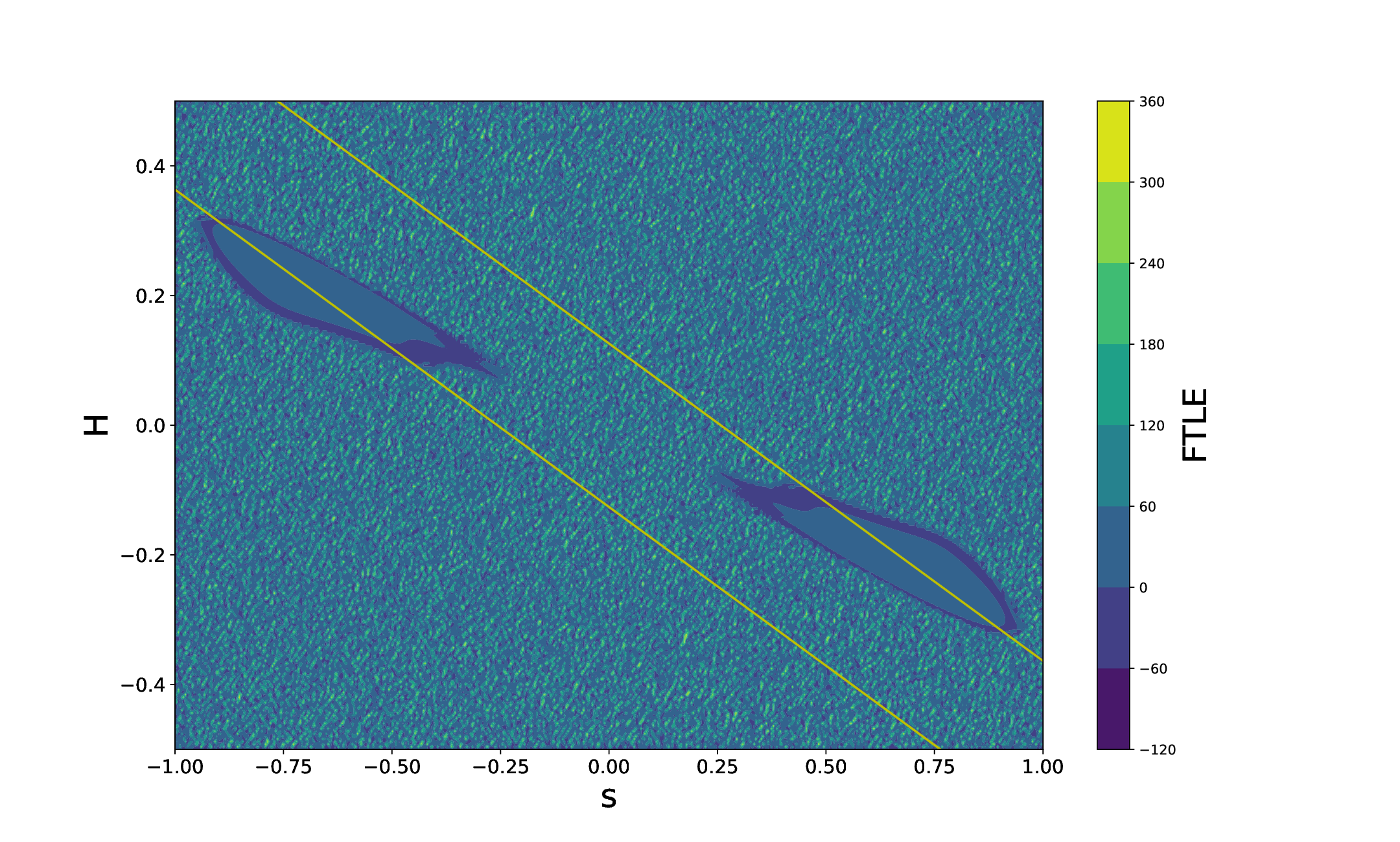}
    (b) 
\end{minipage}
\caption{(a)The discrete,normal Lyapunov exponent $\sigma(s,H)$ is calculated at every point in phase space for $\gamma=0.7$. Phase space values($s,H$) such that $\sigma(s,H)>0$ are marked in red and blue otherwise.(b) The boundaries of the unstable region of phase space are marked as yellow lines on the FTLE plot.}
\label{fig:traj1}
\end{figure*}

With this background, we can now investigate the fate of individual trajectories for several bailout parameter values, starting with $\gamma=10$. Taking a point within the basin of attraction of the periodic attractor($s=-0.75, H=0.2$) with a small perturbation($u=0.001$,$v=0.001$) chosen such that the perturbations are small enough not to cause trajectories to be dislodged from one basin of attraction to another. The dynamics evolves as seen in Fig. \ref{fig:traj}(a)(left), with the map converging to the periodic attractor. Further, the the magnitude of perturbation in the ($u-v$) space computed as $Z=\sqrt{u^2+v^2}$ decays in just a few iterations because of the absence of unstable points on the invariant manifold for $\gamma=10$, as seen in the corresponding plot of $Z$ against the iteration number $n$ in Fig. \ref{fig:traj}(a)(right). Retaining the same value for $\gamma$ and picking a point outside the basin of attraction of the periodic attractor ensures convergence to the chaotic attractor as seen in Fig\ref{fig:traj}(b)(left) for $(s=-0.75, H=-0.2)$. The corresponding perturbation magnitude is seen to decay in just a few iterations of the map, Fig\ref{fig:traj}(b)(right) for the same reason as the earlier initial condition. Increasing the value of $e^{-\gamma}$ by setting $\gamma=1$, we now have a qualitative change in the chaotic attractor with the emergence of an unstable subset as seen earlier in Fig.\ref{fig:nile_plots}(b). The result of this is a transient increase in the magnitude of the perturbation($Z$) for initial conditions on the unstable set which then decays when the trajectory evolves off the unstable subset and onto a stable part of the chaotic attractor. The trajectory of the map and the corresponding transient increase and eventual decrease of the perturbation magnitude are seen in Fig.\ref{fig:traj}(c). It is to be noted here that upon the complete decay of $Z$, dynamics is confined to the ($s, H, u=0,v=0$) plane where the idea of normal instability does not matter and the map can freely explore any area of the chaotic attractor unless a new, non-zero perturbation is introduced in the $(u-v)$ space.
\newline
The final qualitatively distinct picture that arises from the normal instability computations occurs for $\gamma\approx 0.83$ and can be seen in Fig. \ref{fig:nile_plots}(c) where the periodic attractor is unstable. One would expect that this loss of stability would ensure the map converges to the chaotic attractor irrespective of initial condition. However, we see that this is not the case in Fig. \ref{fig:traj}(d) where a trajectory starting very close to the unstable periodic point($s=-0.7711, H=0.2570$) eventually converges to the periodic attractor. While it is the true that the periodic attractor is now normally unstable and results in any arbitrary perturbation growing temporarily( seen in the corresponding $Z$ vs $n$ plot as a transient increase marked by a red bubble), the perturbation decays completely upon leaving the vicinity of the unstable periodic point. With the perturbation having decayed completely($Z=0$), the dynamics is now confined to the invariant manifold where it is free to converge to the periodic attractor. To further understand this we compute the normal Lyapunov exponent($\sigma$) defined in (\ref{eq:3}) once again, but unlike earlier computations of $\sigma$ we no longer evolve the grid of phase space points towards their asymptotic attractors before performing the computation. The result is seen in Fig. \ref{fig:traj1}(a) where we use a similar threshold on the numerical values of $Z$ as used in Fig.\ref{fig:nile_plots}. We observe a diagonal unstable region(red) surrounded by a sea of stability(blue). This computation is particularly insightful when superimposed upon the FTLE plot, where the upper and lower boundaries of the unstable region shown as yellow lines in Fig. \ref{fig:traj1}(b). This reveals the existence of a portion of the periodic attractor's basin of attraction located outside the unstable subset's boundaries. A perturbed trajectory will therefore be able bounce from point to point within the boundaries of the unstable subset until it reaches a point in the stable subset, while still being within the basin of attraction of the periodic attractor. Having reached a stable point within the basin of attraction of the periodic attractor, the disturbance $Z$ decays completely and map converges to towards a point on the periodic attractor. This also justifies our use of the FTLE field for the $Z=0$ case to locate basins of attraction even for regimes with unstable sets on the attractors.

\subsection{Dynamics in the presence of sustained noise}
We have so far restricted ourselves to the study of dynamics in the presence of an isolated perturbation. It is known from literature that the dynamics of bailout embeddings in presence of sustained perturbations may be considerably different from the unperturbed case\cite{Cartwright2002,Cartwright2003}. Indeed, in realistic market scenarios one can expect a sustained perturbation due to noise inherent to financial markets.

An extreme type of intermittency characterized by long periods of relatively quiescent dynamics punctuated by bursts of extreme volatility usually accompanies a complete blow-out bifurcation of the chaotic attractor\cite{Ott1994_1}. This type of dynamics is called on-off intermittency and occurs when the attractor becomes locally unstable in the transverse direction at every point\citep{Ott1994_1,Ashwin1994,Cartwright2003,Ott1996}. In previous sections, blow-out bifurcations have already been demonstrated not to occur in the system considered. However, this does not exclude the possibility of intermittent dynamics which can still occur as long as attractor bubbling and noise are present in the system. To see this, we include additive noise into the bailout embedded map (\ref{eq:2}) such that it serves as repeated perturbation to dynamics on the invariant manifold. The resulting set of equations are thus,

\begin{align}\label{eq:5}
    & s(n+1)=u(n)+\tanh(a\cdot s(n)+b\cdot H(n))\\
    & H(n+1)= v(n)+\theta\cdot H(n)+(1-\theta)s(n)\nonumber\\
    & u(n+1)=e^{-\gamma}[a\sech^2(a\cdot s(n)+b\cdot H(n))u(n) \ldots\nonumber\\
    & +b\sech^2(a\cdot s(n)+b\cdot H(n))v(n)]+\xi\nonumber\\
    & v(n+1)=e^{-\gamma}[(1-\theta)u(n)+\theta v(n)]+\eta\nonumber.
\end{align}
Where $\xi\sim\mathcal{N}(0,10^{-3}),\eta\sim\mathcal{N}(0,10^{-3})$ are independent random variables. Similar to the case of an isolated perturbation, the magnitude of perturbations chosen here are small enough to not change the dynamics qualitatively.
\begin{figure*}

\begin{minipage}{0.4\linewidth}
    \centering
    \includegraphics[width=\linewidth]{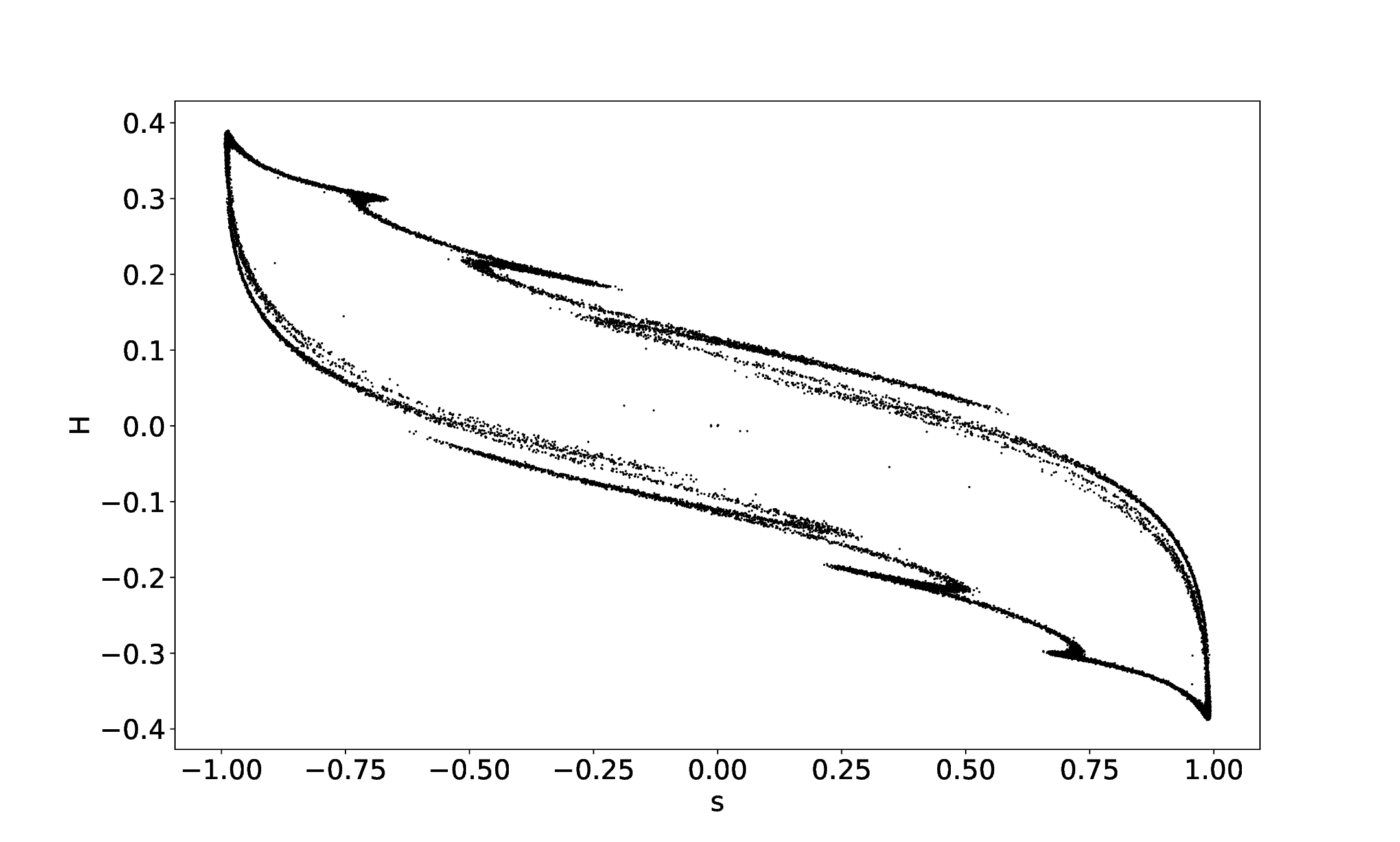}
    (a)
\end{minipage}%
\hspace{0.1\linewidth} 
\begin{minipage}{0.4\linewidth}
    \centering
    \includegraphics[width=\linewidth]{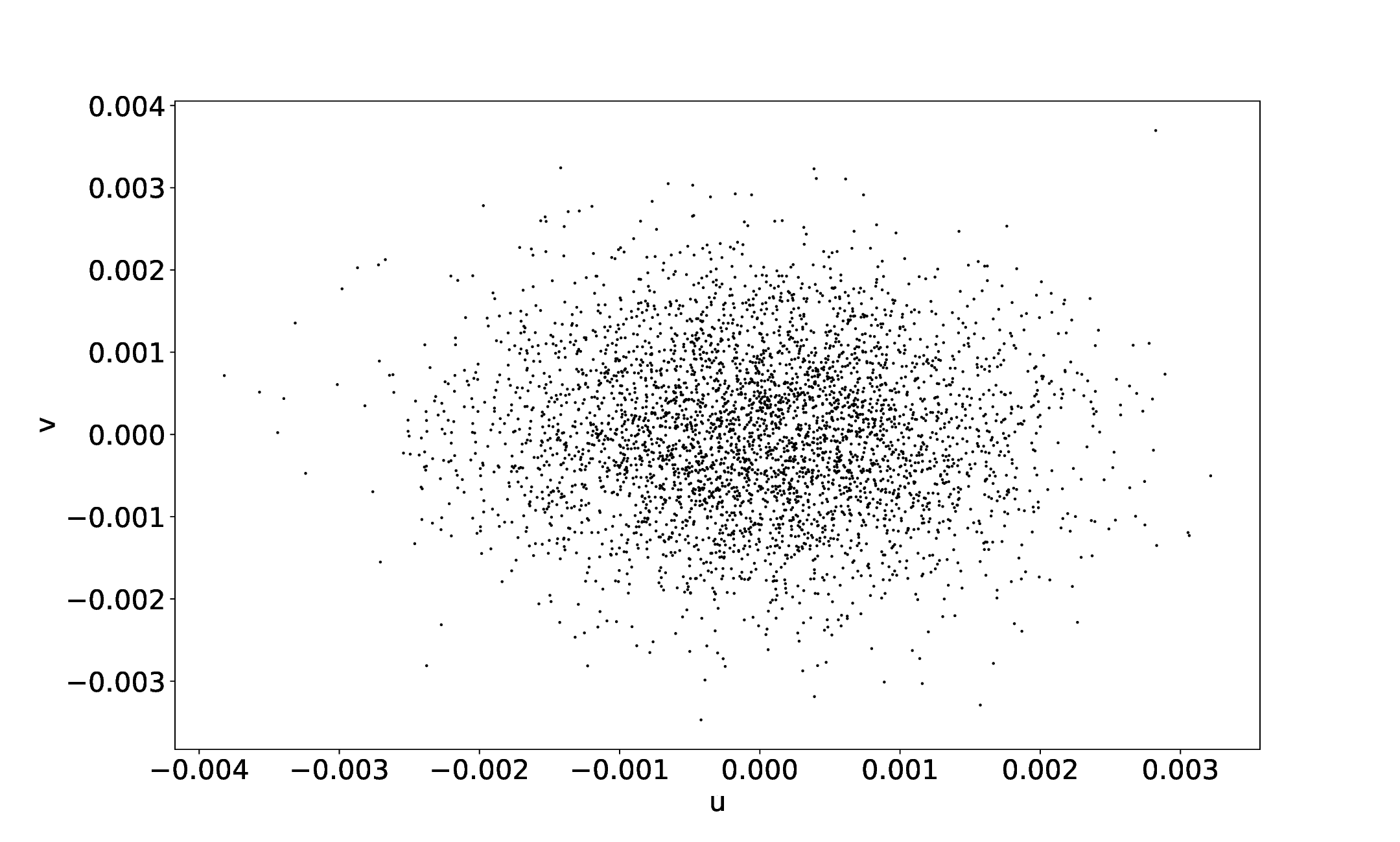}
    (b) 
\end{minipage}

\vspace{0.05\linewidth} 

\begin{minipage}{0.4\linewidth}
    \centering
    \includegraphics[width=\linewidth]{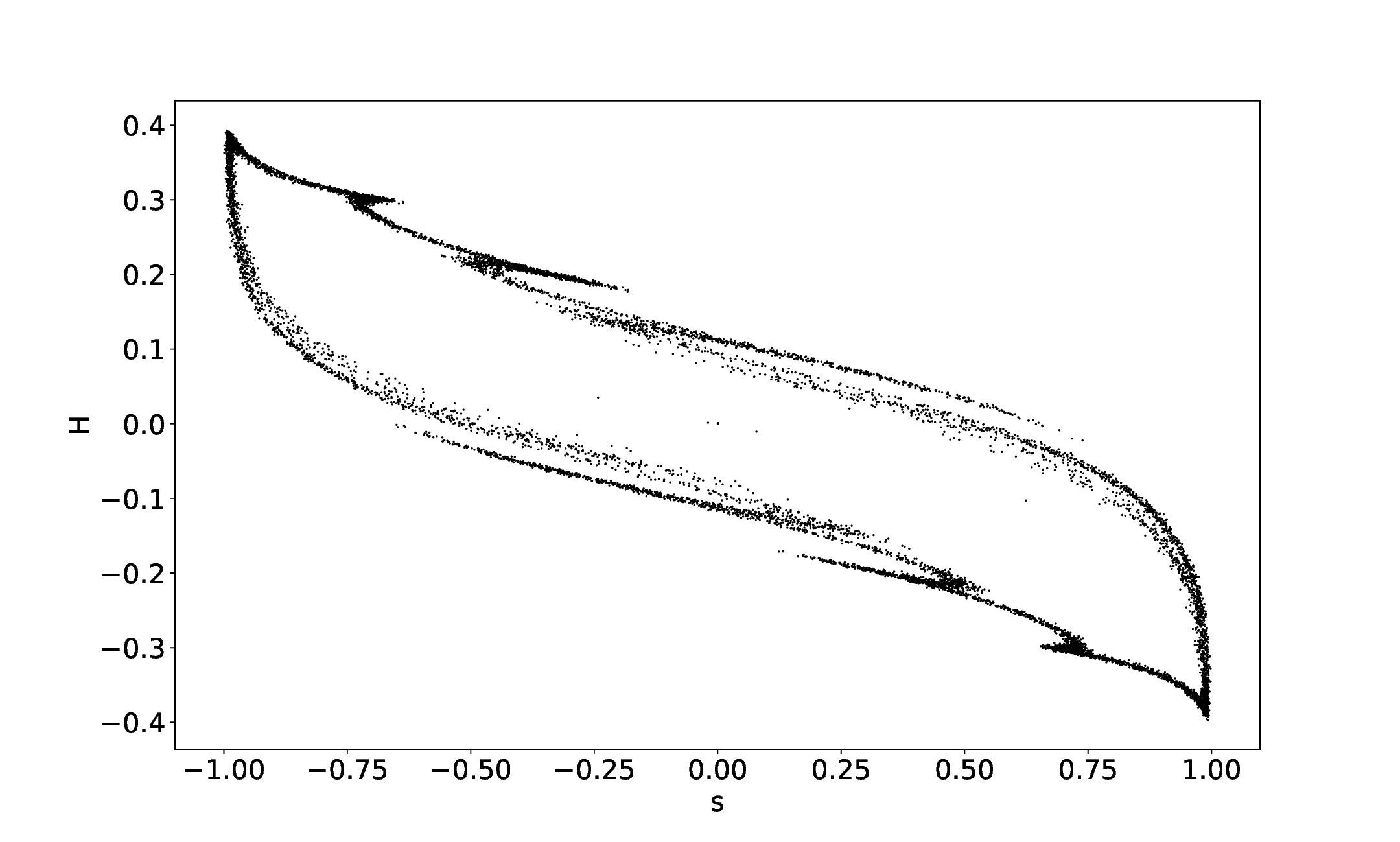}
    (c)
\end{minipage}%
\hspace{0.1\linewidth} 
\begin{minipage}{0.4\linewidth}
    \centering
    \includegraphics[width=\linewidth]{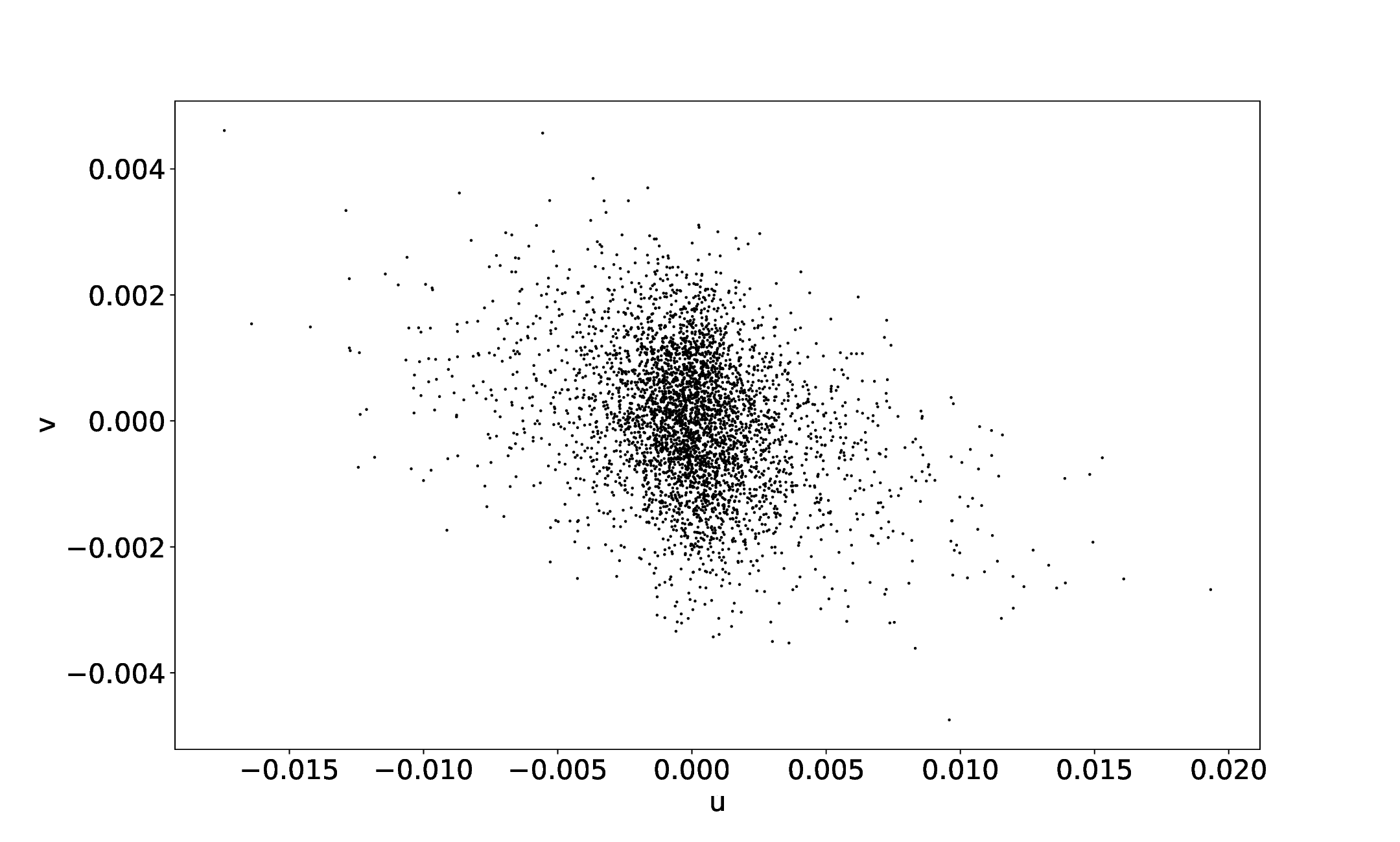}
    (d) 
\end{minipage}

\caption{(a) The trajectory of the map for $\gamma=10$ is identical to the unperturbed dynamics.(b) The dynamics in the $u-v$ space reflects the nature of perturbation chosen.(c) The trajectory of the map for $\gamma=1$ is similar to the $\gamma=10$ case.(d) The dynamics in the $u-v$ space shows larger excursions of the trajectory away from $u=0,v=0$.}
\label{fig:noise1}
\end{figure*}

\begin{figure*}
\begin{minipage}{0.4\linewidth}
    \centering
    \includegraphics[width=\linewidth]{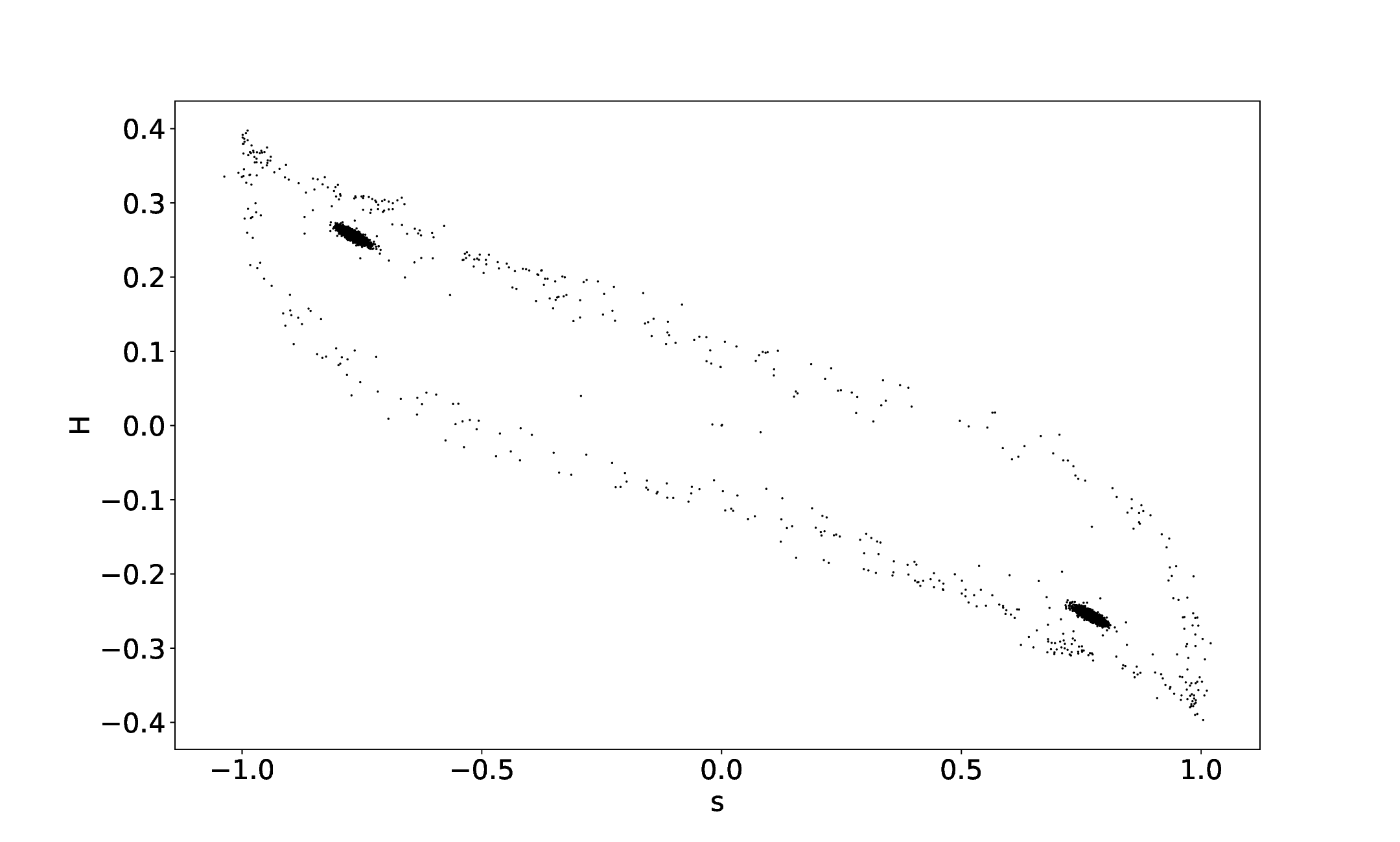}
    (a)
\end{minipage}%
\hspace{0.1\linewidth} 
\begin{minipage}{0.4\linewidth}
    \centering
    \includegraphics[width=\linewidth]{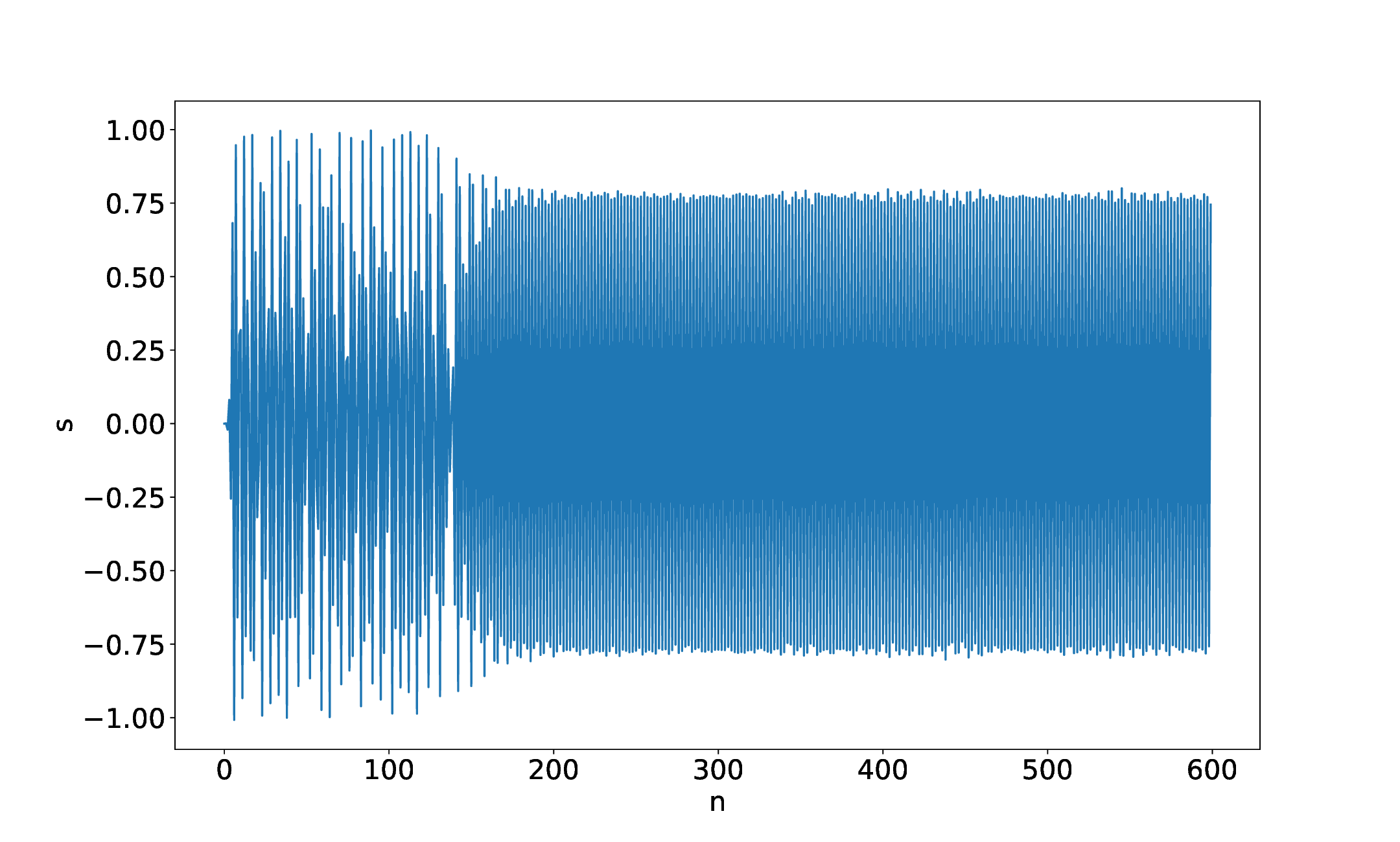}
    (b)
\end{minipage}    

\caption{(a) Dynamics in phase space switches between the chaotic attractor and noisy orbits around the periodic attractor before settling around the periodic points.(b) The discrete time series of the opinion $s$, reflects the phase space dynamics. }
\label{fig:noise2}
\end{figure*}

\begin{figure*}
\begin{minipage}{0.4\linewidth}
    \centering
    \includegraphics[width=\linewidth]{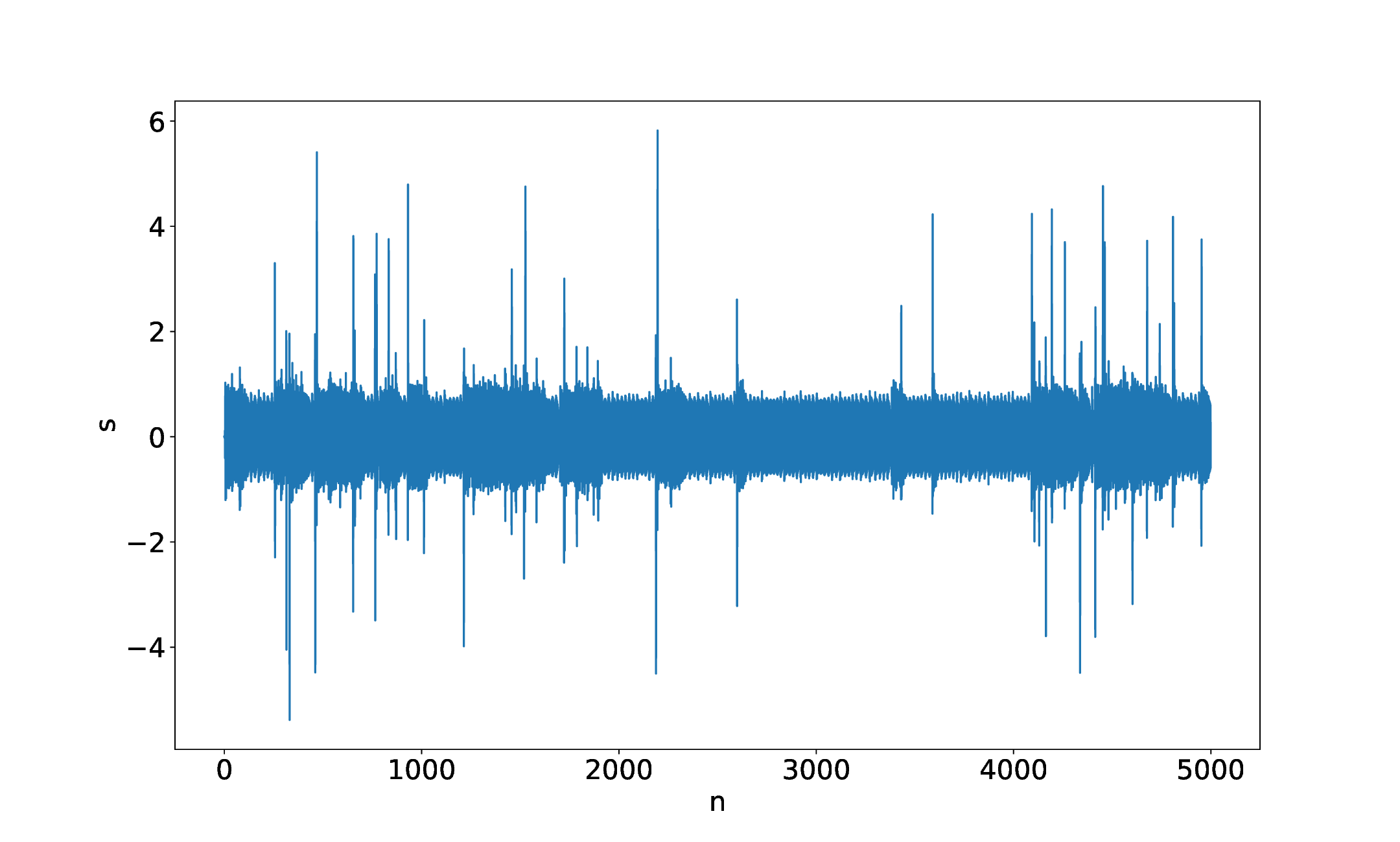}
    (a)
\end{minipage}%
\hspace{0.1\linewidth} 
\begin{minipage}{0.4\linewidth}
    \centering
    \includegraphics[width=\linewidth]{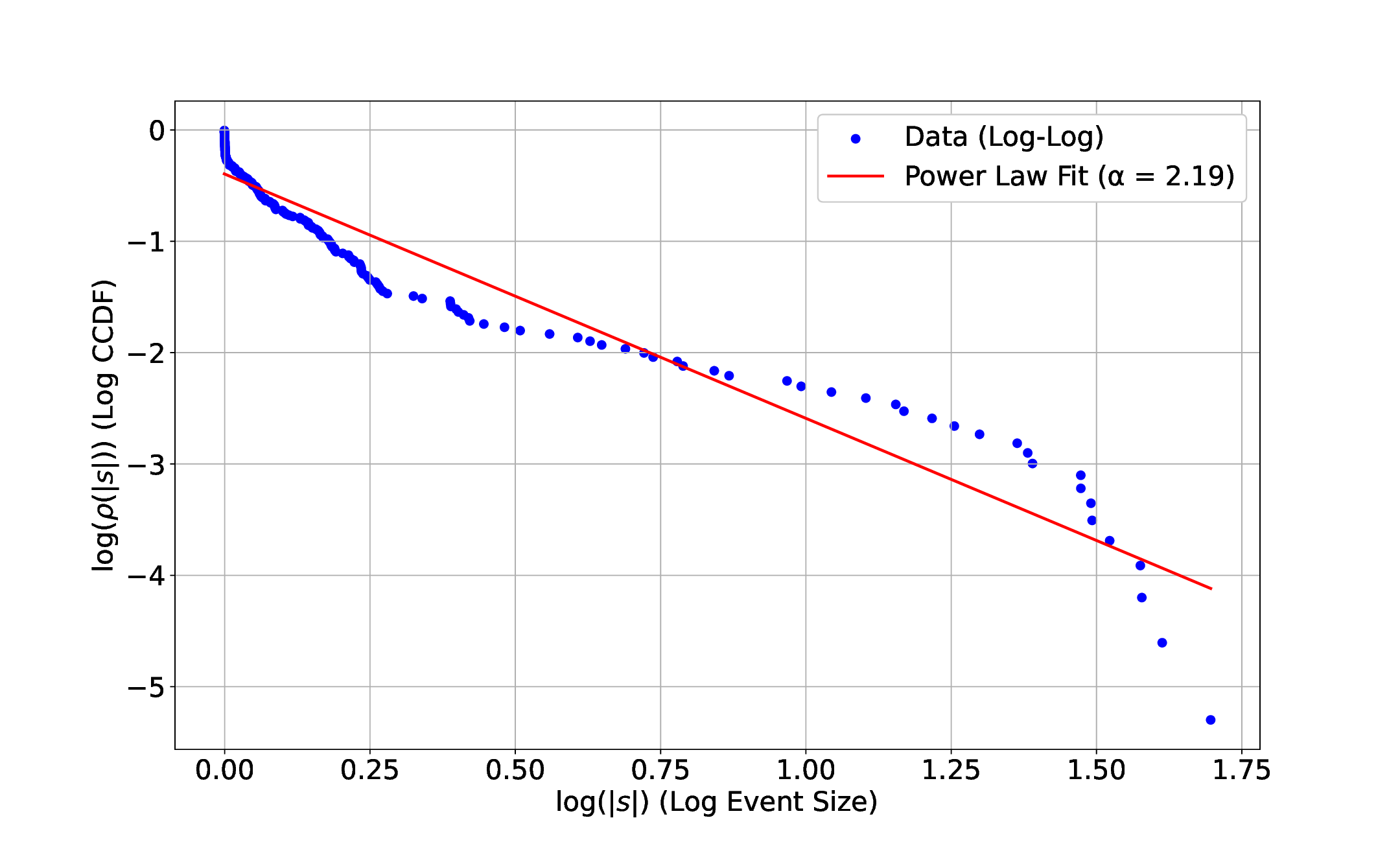}
    (b)
\end{minipage}    
\caption{(a) For $\gamma=0.1$, we plot $s$ over a large number of time steps ($n$).(b) A log-log plot of $|s|$ against the complementary cumulative distribution (CCDF) of $|s|$ (i.e) $log(|s|)$ vs $log(\rho(|s|))$ reveals a power law fit with an exponent of $\alpha\approx 2$. }
\label{fig:noise3}
\end{figure*}

For small values of bailout function(with $\gamma=10$), the attractor is fully stable and any perturbations die out immediately. This causes the dynamics of the map (\ref{eq:5}) to be qualitatively identical to the corresponding case without sustained perturbation as seen in Fig.\ref{fig:noise1}(a). The normally distributed nature of perturbation in $(u-v)$ space, coming from noise is seen in the Fig.\ref{fig:noise1}(b). Upon increasing the value of $e^{-\gamma}$ by setting $\gamma=1$, the unstable subsets of the attractor emerge but the presence of a large stable subset ensures that the dynamics on the attractor remains robust. The only observed effect of the larger bailout function is that the trajectory now traces out a fuzzier version of the chaotic attractor which is difficult to observe from the phase-space plot as seen in Fig.\ref{fig:noise1}(c), but is apparent in ($u-v$) space and is seen as larger excursions of the trajectory as a result of interactions with the unstable subset on the invariant manifold Fig.\ref{fig:noise1}(d). It is to be noted here that the periodic attractor is still stable but it is difficult for the dynamics to balance precisely on the periodic attractor in the presence of sustained noise. However, the system can exhibit noisy orbits around the periodic attractor if noise pushes a trajectory into the low FTLE region surrounding the periodic attractor seen in Fig. \ref{fig:FTLE}. This effect becomes particularly pronounced for larger values of the bailout function where the unstable subset of the chaotic attractor becomes more influential. A particularly interesting case occurs for $\gamma\approx 0.5$, where the trajectory switches back and forth between the noisy orbits surrounding the periodic attractor and chaotic orbits on the chaotic attractor before settling into orbits around the periodic attractor as seen in Fig.\ref{fig:noise2}(a). This is seen as intermittent opinion dynamics for a finite number of discrete time steps followed by periodic dynamics in Fig.\ref{fig:noise2}(b).  With further increase in the value of the bailout function, sudden bursts of large amplitude events emerge in the opinion dynamics for $\gamma\approx 0.1$ as seen in Fig.\ref{fig:noise3}(a) where the sudden bursts of volatility followed by regular dynamics is clearly seen and indicates a type of noise induced intermittency. Thus, we see that perturbation even in the form of simple additive noise can cause the system to exhibit intermittent dynamics with large amplitude events. This differs from some previous demonstrations of intermittency in opinion dynamics that included additive noise as perturbation but also required a multiplicative noise component for the perturbed dynamics to grow and exhibit large deviations away from the invariant manifold\citep{Krawiecki2002}. This observation can be explained by the fact that any disturbance continues growing as long as the trajectory stays within the unstable subset, thus a large unstable subset for high values of the bailout function increases the likelihood that the disturbance attains a large magnitude before the trajectory encounters a stable point. At this juncture, it has to be clarified that unlike on-off intermittency that occurs after a blow-out bifurcation of the invariant manifold, there is a possibility in this system that an arbitrary perturbation starts decaying immediately as long as the current state is located within the stable subset of phase space, regardless of the value of the bailout function. A large value of the bailout function $e^{-\gamma}$ however causes multiple nearby but dis-joint subsets of the attractor to become unstable simultaneously, thereby causing the opinion dynamics to display large jumps that are clustered together in time. This shows that the current state occupied by the dynamics within phase space plays an important role in determining the robustness to noise.

Finally, we study the distribution of extreme events of the intermittent dynamics by plotting the log-log plot of the complementary cumulative distribution function, $\rho(|s|)$ against the event size, $|s|$. Here, we see that the dynamics indeed follows the familiar power law distribution\citep{cont2001}, but with heavier tails than expected. While empirically observed financial returns are known to exhibit power law distribution with exponent closer to $\alpha\approx 3$\citep{Gopikrishnan1998}, this model displays heavier tails with $\alpha\approx 2$ as seen in Fig. \ref{fig:noise3}(b), suggesting the presence of other stabilizing mechanisms in the market that this model does not capture. An interesting extension of this analysis could include the possibility of unstable sets on the invariant manifold that are globally stable despite being completely locally unstable\citep{Tallapragada2017}, which is known to occur in the dynamics of inertial particles\citep{Sudarsanam2017} and could serve as the stabilizing mechanism that is required to obtain more realistic power law exponents.
\section{Bailout function as a measure of investor inertia}\label{section:bof}
The bailout embedding method was developed as a technique to study dynamical systems with lower-dimensional invariant manifolds containing locally unstable subsets and was inspired by the dynamics of inertial particles modeled using the Maxey-Riley equations\citep{MR1983,Cartwright2010}. Under some assumptions and simplifications, the Maxey-Riley equation allows the dynamics of inertial particles in $n$ space dimensions to be modeled as a $2n$ dimensional dynamical system with the original $n$ dimensional subset of phase space acting as an invariant manifold for the dynamics\citep{Babiano2000, Haller2008}. Perturbed particle trajectories on the invariant manifold quickly re-align with the underlying vector field unless the trajectory happens to be passing through an unstable subset of the invariant manifold, in which case it causes the trajectory to detach from the invariant manifold and move into extended, high-dimensional space\cite{Haller2008}. The extent of unstable sets on the invariant manifold depends on the size of the particle, with larger subsets of the invariant manifold being unstable to the trajectories of larger particles(higher inertia) and smaller unstable sets for the dynamics of smaller sized particles\cite{Babiano2000}. In bailout embeddings of the Maxey-Riley equation\citep{Cartwright2002}, inertia of the particle is captured within the exponent $\gamma$ in the bailout function $K(x)=e^{-\gamma}\frac{\partial f}{\partial x}$ such that $\gamma$ is inversely proportional to the size of the inertial particle. Thus, for large particle inertia we have small values of $\gamma$ and relatively large values of the bailout function $e^{-\gamma}$ as a consequence, which in turn determines the extent of unstable sets on the invariant manifold. 

Inspired by the role played by the bailout function in the context of inertial particle dynamics, one can envision the bailout function in opinion dynamics as a measure of investor inertia\citep{Cont_bouchaud2000}. In this context, investor inertia can be interpreted as the reluctance of individual agents to change their prevailing opinion. In other words, the bailout function can be understood as a measure of collective slowing down of opinion dynamics. By steadily increasing the value of the bailout function, we go from a low bailout function value regime where the individual agents are nimble and are able to quickly re-adjust their opinions upon perturbation, to a high bailout function value regime where there is larger inertia in the response of market participants and the consequent inability to recover from perturbations quickly. Thus, by varying the bailout parameter we artificially induce higher inertia into the dynamics which in turn causes market crash-like extreme events. This ability to induce high investor inertia by tuning the bailout function enables us to study the effect of investor inertia under a controlled setting.
\newline
We know from earlier analysis that all subsets of state space do not react uniformly to a change in investor inertia as measured by the bailout function. This is similar to the case of inertial particle dynamics where subsets of the flow field dominated by deformation are more unstable than regions dominated by rotation\citep{Babiano2000}. The stability analysis from section \ref{section:stability} and the corresponding results seen in Figs.\ref{fig:nile_plots}(a-d) point to the fact that instability manifests itself most strongly in market regimes where the current opinion $s$ of a representative agent is different in polarity(sign) from the recent trend($H$) obtained by averaging the previous $\frac{1}{1-\theta}$ time steps in the opinion states. This can be understood as the situation where the opinion of a representative market participant has been trending positive or negative for a finite duration but they now hold a contrary opinion. While our choice of negative values for the parameters $a,b$ in model (\ref{eq:1}) has introduced deliberate contrarian behavior among the agents in the model, incorporation of the new value of $s$ into the trend $H$ is slowed down for values of $e^{-\gamma}>0$, thus leading to instabilities on the attractor.  These results underline the importance of adaptability in the markets and the ability to react quickly to emerging news becomes critical in preventing the onset of an endogenously induced crash.
\section{Conclusion}\label{section:conclusion}
In this work we study a two-dimensional, discrete-time model of opinion dynamics in the financial markets. With the goal of studying the stability of attractors in the dynamics, we focus our attention to a particular parameter regime that is known to have multiple co-existing attractors in phase space. We employ the bailout embedding technique to embed the discrete-time system in a higher-dimensional space such that the original dynamics is preserved on an invariant manifold in two space dimensions while the trajectory of the system detaches from the invariant manifold if perturbed in a direction orthogonal to the invariant manifold. We demonstrate the various stability regimes that the underlying periodic and chaotic attractors undergo as the bailout parameter is varied. We demonstrate the qualitatively different trajectories that are exhibited when initial conditions are chosen in different subsets of phase space separated by ridges of the Finite Time Lyapunov Exponent(FTLE) field. We also demonstrate the decay or transient growth of magnitude ($Z$) of a perturbation in the higher-dimensional disturbance space, for various stability regimes. Finally, we demonstrate the occurrence of intermittency in opinion dynamics when perturbed by sustained additive noise at a sufficiently high value of the bailout function. Inspired by mathematical models of finite sized particle dynamics in fluids, we hypothesize that the bailout function resembles a measure of inertia in opinion dynamics within financial markets.
We observe that the 2-D model used to study opinion dynamics in this work avoids a complete blow-out bifurcation of the invariant manifold due to the presence of multiple attractors with different levels of stability. This suggests a possible mechanism by which markets may exhibit stable dynamics even in large inertia regimes. While the contribution of mechanisms such as attractor bubbling and blow-out bifurcations to intermittent behavior in financial opinion dynamics is fairly well established, we conduct a thorough systematic examination of the stability of various subsets on the invariant manifold for a particular case of a more general class of mean-field Ising models. Despite the restricted parameter range of this investigation, the examination reveals several insights. First, we glimpse the possibility for emergence of highly clustered extreme events even without multiplicative noise forcing and in the absence of a complete blow-out bifurcation of the invariant manifold. Second, this model enables us to undertake a systematic study of the implications of increasing investor inertia. Third, the investigation hints at a novel mechanism for market stabilization in high investor inertia regimes via locally unstable yet globally stable subsets of attractors. We obtain these insights using the bailout embedding technique which we propose here as a natural tool for the systematic stability analysis of financial dynamical systems with sustained noise perturbation.
\section*{Author Declarations}
The authors have no conflicts to disclose.
\section*{Data Availability}
The data that support the findings of this study are available from the corresponding author upon reasonable request.
\section*{Appendix}
The bailout embedding of a map $x_{n+1}=f(x_n)$ is a map of the form $x_{n+2}-f(x_{n+1})=K(x_n)[x_{n+1}-f(x_n)]$ with $K(x_n)=e^{-\gamma}\frac{\partial f}{\partial x_n}$. The Bailout embedding of the 2D map (\ref{eq:1}) is as follows:
\setcounter{equation}{0}
\begin{equation}
X_{n+2}-f(X_{n+1})=K(X_n)[X_{n+1}-f(X_{n})]    
\end{equation}

Where $X_n=\bigg[\begin{array}{cc}
     & s_n \\
     & H_n
     \end{array}\bigg]$
is a 2-D vector and $f(X_n)=\bigg[\begin{array}{cc}
     & f_s(s_n,H_n) \\
     & f_H(s_n,H_n)
     \end{array}\bigg]$
with $f_s(s_{n},H_{n})=\tanh(a\cdot s_n+b\cdot H_n)$ and $f_H(s_n,H_n)=\theta\cdot H_n+(1-\theta)s_n$ from the original 2-D system.\vspace{0.5cm}\\   
Introducing a new set of variables $u_n,v_n$, such that, $W_n=\bigg[\begin{array}{cc}
     & u_n \\
     & v_n
     \end{array}\bigg]$ and setting $K(X_n)[X_{n+1}-f(X_{n})]=W_{n+1}$ in (\ref{eq:1}), we get 
\begin{equation*}
X_{n+2}-f(X_{n+1})=W_{n+1}.
\end{equation*}
Alternatively, one can write,
\begin{eqnarray}
    &s_{n+2}-f_s(s_{n+1},H_{n+1})=u_{n+1}\nonumber\\
    &\text{or, }s_{n+1}=u_{n}+f_s(s_{n},H_{n})\\
    &\text{and, }H_{n+2}-f_H(s_{n+1},H_{n+1})=v_{n+1}\nonumber\\
    &\text{or, }H_{n+1}=v_{n}+f_H(s_{n},H_{n})
\end{eqnarray}
by expanding out $W_n$ and writing out equations for $u_n,v_n$ individually.\vspace{0.5cm}\\ 
We already know that $W_{n+1}=K(X_n)[X_{n+1}-f(X_{n})]$, which, by using (\ref{eq:2}-\ref{eq:3}) can be written as,
$\bigg[\begin{array}{cc}
     & u_{n+1} \\
     & v_{n+1}
     \end{array}\bigg]=K(X_n)\bigg[\begin{array}{cc}
     & u_n \\
     & v_n
     \end{array}\bigg]$\\
     \vspace{0.5cm}
where,\\
    $K(X_n)=e^{-\gamma}\bigg[\begin{matrix}
     & a\sech^2(a\cdot s_n+b\cdot H_n) 
     &  b\sech^2(a\cdot s_n+b\cdot H_n) \\
     & 1-\theta 
     & \theta
     \end{matrix}\bigg].$
This yields the following set of equations,
\begin{eqnarray}
    &u_{n+1}=e^{-\gamma}[a\sech^2(a\cdot s_n+b\cdot H_n)u_n\ldots\nonumber\\
    & +b\sech^2(a\cdot s_n+b\cdot H_n)v_n]\\
    &v_{n+1}=e^{-\gamma}[(1-\theta)u_n+\theta v_n].
\end{eqnarray}
The set of four equations (2-5) constitute the 4-D bailout embedding under study in this work.
\clearpage
\bibliography{aipsamp}
\end{document}